\documentclass{aa}
\usepackage{graphicx}
\usepackage{subfigure}
\usepackage{multirow}
\usepackage{amssymb}
\usepackage{natbib}
\usepackage{txfonts}
\usepackage{rotating}
\usepackage[bookmarks=false,colorlinks=true, citecolor=blue]{hyperref}
\begin{document}
 \title{$\rm H_2$ formation and excitation in the Stephan's Quintet galaxy-wide collision}


 \author{P. Guillard\inst{1}
        \and
        F. Boulanger\inst{1}
		  \and
		  G. Pineau des For\^ets\inst{1,3}
		  \and
	      P.N. Appleton\inst{2}
        }

  \authorrunning{P.~Guillard et al.}
  \titlerunning{$\rm H_2 $ formation and excitation in galaxy-wide collisions}
  \offprints{P.~Guillard, \email{pierre.guillard@ias.u-psud.fr}}

 \institute{Institut d'Astrophysique Spatiale (IAS), UMR 8617, CNRS, Universit\'e
  Paris-Sud 11, B\^atiment 121, 91405 Orsay Cedex, France
  \and NASA \emph{Herschel} Science Center (NHSC), California Institute of Technology, Mail code 100-22, Pasadena, CA 91125, USA
 \and  LERMA, UMR 8112, CNRS, Observatoire de Paris, 61 Avenue de l'Observatoire, 75014 Paris, France
 }

\date{Received ? / accepted ? }


\abstract
{The \textit{Spitzer Space Telescope} has detected a powerful ($L_{\rm H_2}\sim10^{41}$~erg~s$^{-1}$) mid-infrared H$_{2}$ emission towards the galaxy-wide collision in the Stephan's Quintet (SQ) galaxy group. This discovery was followed by the detection of more distant H$_{2}$-luminous extragalactic sources, with almost no spectroscopic signatures of star formation. These observations place molecular gas in a new context where one has to describe its role as a cooling agent of energetic phases of galaxy evolution.
}
 {The SQ postshock medium is observed to be multiphase, with H$_{2}$ gas coexisting with a hot ($\sim5\times10^{6}$~K), X-ray~emitting~plasma. The surface brightness of H$_{2}$ lines exceeds that of the X-rays and the 0-0~S(1)~$\rm H_2$ linewidth is $\sim900$~km~s$^{-1}$, of the order of the collision velocity. These observations raise three questions we propose to answer: \textit{(i)} Why~H$_{2}$ is present in the postshock gas~? 
\textit{(ii)}~How can we account for the $\rm H_2$ excitation~? \textit{(iii)}~Why is H$_2$ a dominant coolant~?}
 {We consider the collision of two flows of multiphase dusty gas. Our model quantifies the gas cooling, dust destruction, H$_{2}$ formation and excitation in the postshock medium.}
 {
\textit{(i)} The~shock~velocity, the post-shock temperature~and~the~gas~cooling timescale depend on the preshock gas density. The collision velocity is the shock velocity in the low density volume filling intercloud gas. This produces a $\sim5\times10^6~$K, dust-free, X-ray~emitting~plasma. The shock velocity is smaller in clouds. We show that gas heated to temperatures less than $10^6~$K cools, keeps its dust content and becomes H$_2$ within the SQ collision age ($\sim5\times10^6~$years). \textit{(ii)} Since the bulk kinetic energy of the H$_2$ gas is the dominant energy reservoir, we consider that the H$_2$ emission is powered by the dissipation of kinetic turbulent energy. We model this dissipation with non-dissociative MHD shocks and show that the H$_2$ excitation can be reproduced by a combination of low velocities (in the range $5 - 20$~km~s$^{-1}$) shocks within dense
($n_H>10^3 ~cm^{-3}$) H$_2$ gas. 
\textit{(iii)} An~efficient~transfer~of~the bulk kinetic energy to turbulent motions of much lower velocities within molecular gas is required to make H$_2$ a dominant coolant of the postshock gas. We argue that this transfer is mediated by the dynamical interaction between gas phases and the thermal instability
of the cooling gas. We~quantify~the~mass~and~energy~cycling~between~gas~phases~required~to~balance~the~dissipation~of~energy through~the~H$_2$ emission lines. 
}
{This study provides a physical framework to interpret H$_2$~emission from H$_2$-luminous~galaxies. It highlights the role that H$_2$~formation and cooling
play in dissipating mechanical energy released in galaxy collisions. This physical framework is of general relevance for the interpretation of observational signatures, in particular H$_2$~emission,
of mechanical~energy~dissipation~in multiphase gas.}


\maketitle


\section{Introduction}

\emph{Spitzer} Infra-Red Spectrograph (IRS) observations of the
Stephan's Quintet (hereafter SQ) galaxy-wide collision led to the
unexpected detection of extremely bright mid-IR $\rm H_2$ rotational
line emission from warm ($\sim 10^{2} - 10^{3}\,$K) molecular gas
\citep{2006ApJ...639L..51A}. This is the first time an almost pure
$\rm H_2$ line spectrum has been seen in an extragalactic object. This
result is surprising since $\rm H_2$ is coexisting with a hot X-ray
emitting plasma and almost no spectroscopic signature (dust or ionized
gas lines) of star formation is associated with the $\rm H_2$
emission. This is unlike what is observed in star forming galaxies
where the $\rm H_2$ lines are much weaker than the mid-IR dust
features \citep{Rigopoulou2002, Higdon2006, 2007ApJ...669..959R}.

The detection of $\rm H_2 $ emission towards the SQ shock was quickly
followed by the discovery of a class of $\rm H_2$-luminous galaxies
which show high equivalent width $\rm H_2 $ mid-IR lines.  The first
$\rm H_2$-luminous galaxy, NGC~6240, was identified from ground based
near-IR $\rm H_2 $ spectroscopy by \citet{Joseph1984}, although this
galaxy also exhibits a very strong IR continuum.  Many more of $\rm
H_2$ bright galaxies are being found with Spitzer \citep{Egami2006,
  Ogle2007}, often with very weak continuua suggesting that star formation
is not the source of $\rm H_2 $ excitation. These $\rm H_2$-luminous
galaxies open a new perspective on the role of $\rm H_2$ as a cold gas
coolant, on the relation between molecular gas and star formation, and
on the energetics of galaxy formation, which has yet to be explored.

SQ is a nearby $\rm H_2$-luminous source where observations provide an
unambigious link between the origin of the $\rm H_2$ emission, and a
large-scale high-speed shock. The SQ galaxy-wide shock is created
by an intruding galaxy colliding with a tidal tail at a relative
velocity of $\sim 1\,000$~km~s$^{-1}$.  Evidence for a group-wide
shock comes from observations of X-rays from the hot postshock gas in the ridge 
\citep{2003A&A...401..173T, 2005A&A...444..697T, O'Sullivan2008},
strong radio synchrotron emission from the radio emitting plasma
\citep{2001AJ....122.2993S} and shocked-gas excitation diagnostics
from optical emission lines \citep{2003ApJ...595..665X}.
\textit{Spitzer} observations show that this gas also contains
molecular hydrogen and that it is turbulent with an $\rm H_2$
linewidth of 870 km~s$^{-1}$. The $\rm H_2$ surface brightness is
larger than the X-ray emission from the same region, thus the $\rm
H_2$ line emission dominates over X-ray cooling in the shock. As such,
it plays a major role in the energy dissipation and evolution of the
postshock gas. 

These observations raise three questions we propose to answer: \textit{(i) Why is
there $ \rm H_2 $ in the postshock gas~? (ii) How can we account for the $ \rm H_2 $ excitation~? (iii) Why is H$_2$ a dominant coolant~?} We introduce these three questions, which structure this paper.

\textit{(i) H$_2$ formation.} The detection of large quantities of warm molecular gas in  the SQ shock, coexisting with a X-ray emitting plasma, is a surprising result. \citet{2006ApJ...639L..51A} invoked an oblique shock geometry, which would reduce the shock strength. However, with this hypothesis, it is hard to explain why the temperature of the postshock plasma is so high ($5 \times 10^{6}$~K, which contrains the shock velocity to be $\sim 600$~km~s$^{-1}$).
Therefore the presence of H$_2$ is likely to be related to the multiphase, cloudy structure of the preshock gas. 

One possibility is that molecular clouds were present in the preshock gas. 
Even if the transmitted shock into the cloud is dissociative, $\rm H_2$ molecules may reform in the postshock medium \citep{1979ApJS...41..555H}. An alternative possibility is that $\rm H_2$ forms out of preshock H{\sc i} clouds.
 \citet{2006ApJ...639L..51A} proposed that a large-scale shock
overruns a clumpy preshock medium and that the $\rm H_2$ would form in
the clouds that experience slower shocks.
In this paper we quantify this last scenario by considering the collision between two inhomogeneous gas flows, one being associated with the tidal tail and the other associated with the interstellar medium (hereafter ISM) of the intruding galaxy. 

\textit{(ii) H$_2$ excitation.}  
Several excitation mecanisms may account for high-equivalent width $\rm H_2$ line emission. 
The excitation by X-ray photons was quantified by a number of authors  \citep[e.g.][]{Tine1997, Dalgarno1999}. The energy conversion from X-ray flux to $\rm H_2$ emission is at most 10\% for a cloud that is optically thick to X-ray photons. The absorbed fraction of X-ray photons may be even smaller if the postshock $\rm H_2$ surface filling factor is smaller than 1.
\textit{Chandra} and \textit{XMM} observations ($0.2 - 3$~keV) show that, within the region where \textit{Spitzer} detected  $\rm H_2$ emission, the $\rm H_2$ to X-ray luminosity ratio is $\sim 5$ \citep{O'Sullivan2008, 2005A&A...444..697T}. Therefore, the excitation of $\rm H_2$ by $0.2 - 3$~keV X-ray photons cannot be the dominant process. 

$\rm H_2$ excitation may also be produced by cosmic ray ionization \citep{Ferland2008}. However radio continuum observations of SQ show that the combined cosmic ray plus magnetic energy is not the dominant energy reservoir.

SQ observations suggest that only a fraction of the collision energy is used to heat the hot plasma. Most of this energy is observed to be kinetic energy of the $\rm H_2$ gas. 
Therefore we consider that the H$_2$ emission is most likely to be powered by the dissipation of the kinetic energy of the gas.
This excitation mechanism has been extensively discussed for the Galactic interstellar medium. It has been proposed to account for H$_2$ emission from solar neighbourhood clouds \citep{Gry2002, Nehm'e2008}, from the diffuse interstellar medium in the inner Galaxy \citep{Falgarone2005} and from clouds in the Galactic center \citep{Rodr'iguez-Fern'andez2001}.

\textit{(iii) H$_2$ cooling.} 
For $\rm H_2$ to be a dominant coolant of the SQ postshock gas, the kinetic energy of the collision in the center of mass rest-frame has to be efficiently transfered to the molecular gas. In the collision, the low density volume filling gas is decelerated but the clouds keep their preshock momentum. The clouds which move at high velocity with respect to the background plasma are subject to drag. This drag corresponds to an exchange of momentum and energy with the background plasma.  
This transfer of energy between clouds and hot plasma has been discussed in diverse astrophysical contexts, in particular infalling clouds in galaxy clusters cooling flows \citep{2008MNRAS.389.1259P} and galactic halos \citep{Murray2004}, as well as  cometary clouds in the local interstellar medium \citep{Nehm'e2008} and the Helix planetary nebula tails \citep{Dyson2006}.

The dynamical evolution of the multiphase interstellar medium (ISM) in galaxies has been extensively investigated with numerical simulations, within the context of the injection of mechanical energy by star formation, in particular supernovae explosions \citep[e.g.][]{2005, Dib2006}. The injection of energy by supernovae is shown to 
be able to maintain a fragmented, multiphase and turbulent ISM. Numerical simulations on smaller scales show that the dynamical interaction between gas phases and the thermal instability feed turbulence within clouds \citep{2003ApJ...591..238S, Audit2005, Nakamura2006}.
The SQ collision releases $\sim 10^{56}$~erg of mechanical energy in a collision timescale of a few million years. This is commensurate with the amount of mechanical energy injected by supernovae in the Milky  Way (MW) over the same timescale. The main difference is in the mass of gas, which is two orders of magnitude smaller in the SQ shock than in the MW  molecular ring. 
None of the previous simulations apply to the context of the SQ collision, but they provide relevant physics to our problem.
In this paper, based on this past work, we discuss how the dynamical interaction between the molecular gas and the background plasma may sustain turbulence within molecular clouds at the required level to balance the dissipation of energy through the H$_2$ emission lines. 



The structure of this paper is based on the previous 3 questions and is organised as follows: in \S~\ref{sec_observations}, we gather the observational
data that set the luminosity, mass and energy budget of the SQ
collision. In \S~\ref{section_scenario}, the gas cooling, dust destruction, and $\rm H_2$ formation in the postshock gas are quantified. In \S~\ref{sec_H2excitation} the dissipation of  kinetic energy within the molecular gas is modeled, in order to account  for the $\rm H_2$ emission in SQ. Section~\ref{sec_cycle} presents a view at the dynamical interaction between ISM phases that arises from our interpretation of the data. In \S~\ref{sec_implications_dissipation} we discuss why $\rm H_2$ is such a dominant coolant in the SQ ridge.  Section~\ref{sec_future_tests} discusses open questions including future observations. Our conclusions are presented in \S~\ref{section_conclusion}.

\section{Observations of Stephan's Quintet}\label{sec_observations}

This section introduces the observational data that set the
luminosity, mass and energy budgets of the SQ shock. Relevant
observational numbers and corresponding references are gathered in Table~\ref{table_budgets}. Note that all the quantities are scaled to the aperture $\mathcal{A}$
used by \citet{2006ApJ...639L..51A} to measure $\rm H_2$
luminosities~:~$\mathcal{A} = 11.3 \times 4.7 ~\rm arcsec^2$, which
corresponds to $5.2 \times 2.1~\rm kpc^{2}$. 
The equivalent volume associated with this area is $\mathcal{V_A} = 5.2 \times 2.1 \times l_{z}~\rm kpc^{3}$, where $l_{z}~\rm [kpc]$ is the dimension along the line-of-sight. For the X-ray emitting gas, \citet{O'Sullivan2008} assume a cylindrical geometry for the shock, which gives $l_{z} \sim 4.5$~kpc.
This dimension may be smaller for the $\rm H_2$ emitting gas, which could be concentrated where the gas density is largest. To take this into account, we use a reference value of $l_z = 2$~kpc. Since this number is uncertain, the explicit dependence of the physical quantities on this dimension are written out in the equations.

\subsection{Astrophysical context}\label{subsec_context}

Since its discovery \citep{Stephan1877}, the Stephan's Quintet 
(94~Mpc) multigalaxy compact group has been extensively studied, 
although one member of the original quintet (NGC 7320) is now presumed to be a foreground dwarf and will not be discussed further here. Later studies
have restored the ``quintet'' status by the discovery of a more distant member. The inner group galaxies are shown in the left-hand side of Fig.~\ref{Fig_picture_shocks}.

H{\sc i} observations exhibit a large stream
created by tidal interactions between NGC~7320c, and NGC~7319 \citep{Moles1997, 2001AJ....122.2993S}. It is postulated that another galaxy, NGC~7318b (hereafter called the ``intruder''), is falling into the group with a very high relative velocity shocking a $25$~kpc segment of that  tidal tail \citep{2003ApJ...595..665X}. The observed difference in radial velocities between the tidal tail and the intruder is $\sim 1\,000$~km~s$^{-1}$. If this picture of two colliding gas flows is right, we expect two shocks: one propagating forward  into the SQ intergalactic medium, and an other reverse shock driven into the intruding galaxy.

Over the shock region, no H{\sc i} gas is
detected, but optical line emission from H{\sc ii} gas is observed at
the H{\sc i} tidal tail velocity. X-ray \citep{2003A&A...401..173T,
  2005A&A...444..697T, O'Sullivan2008} observations resolve shock-heated plasma
spatially associated with the optical line and 21~cm radio continuum
\citep{2001AJ....122.2993S}. The width of this emission is
$\sim 5$~kpc, which is commensurate with the width of the H$\, \alpha$ emission. Dividing this shock width by the collision velocity (assumed to be $1\,000$~km~s$^{-1}$), we get a
collision age of $t_{\rm coll} \sim 5 \times 10^{6} \,$yr.

Spectral energy distribution fitting of \emph{Chandra} and \emph{XMM} observations of SQ show that the temperature of the hot plasma is $T_{\rm X} \sim 5 \times 10^{6} \,$K \citep{2003A&A...401..173T, 2005A&A...444..697T, O'Sullivan2008}.
The postshock temperature is given by \citep[see e.g.][]{1993ARA&A..31..373D}:
\begin{equation}
T_{\rm ps} = \frac{2(\gamma - 1)}{(\gamma + 1)^{2}} \, \frac{\mu}{k_{\rm B}} \, V_{\rm s}^{2} \simeq 5 \times 10^6 \, \left( \frac{V_{\rm s}}{600 \ \rm km \, s^{-1}}\right) ^{2} \ \ \rm K \; ,
\end{equation}
where $V_{\rm s}$ is the velocity of the shock wave, $\mu$ the mean particle mass (equals to $10^{-24}$~g for a fully ionized gas), $k_B$ the Boltzmann constant and $\gamma = 5/3$. Therefore, the postshock temperature of the hot plasma allows to estimate a shock velocity of $\sim 600$~km~s$^{-1}$. This shock velocity is consistent with the gas velocity in the center of mass frame, which would be approximately half of the observed relative velocity 
 if the mass of the gas reservoirs on the intruder and SQ tail sides are commensurate.

\subsection{Mass budgets}\label{subsec_mass_nrj_budget}

\begin{table*}
\begin{center}
\begin{minipage}[t]{18cm}
 \renewcommand{\footnoterule}{}
\def\thefootnote{\alph{footnote}}
\centering
    \begin{tabular}{c || c c c || c c c c c}
	\hline
	\hline
		& \multicolumn{3}{c ||}{Preshock Gas} &  \multicolumn{5}{c}{Postshock Gas} \\
  \hline
	\hline
   & Hot Plasma\footnotemark[1] & H{\sc i}\footnotemark[3] & $\rm H_2$ &  Hot Plasma\footnotemark[1] & H{\sc ii}\footnotemark[2]  & H{\sc i}\footnotemark[3] & Warm $\rm H_2$\footnotemark[4]& Cold $\rm H_2$ \\
  Density $n_{\rm H}$ [cm$^{-3}$] & & & & 0.02 &  & & & \\
  Temperature  $T$ [K] &  &  & & $5 \times 10^{6}$ &  & & 185 & $< 10^{2}$ \\
  Pressure $P / k_{\rm B}$ [ K cm$^{-3}$] &  &  & & $2 \times 10^{5}$ &  &  & &   \\
   $N_{\rm H}$ [cm$^{-2}$] & & $3 \times 10^{20}$ & & & $1.4 \times 10^{19}$  & $< 5.8 \times 10^{19}$ & $2 \times 10^{20}$ & \\
  \hline
	\hline
	\multirow{2}*{Mass Budget [$\rm M_{\odot}$]} & Hot Plasma\footnotemark[1] & H{\sc i}\footnotemark[3] & $\rm H_2$ &  Hot Plasma\footnotemark[1] & H{\sc ii}\footnotemark[2]  & H{\sc i}\footnotemark[3] & Warm $\rm H_2$\footnotemark[4]& Cold $\rm H_2$ \\
     & $ 6 \times 10^6 $ &  $2 - 6 \times 10^{7}$ &  &  $ 1.4 \times 10^{7}$ & $1.2 \times 10^6$ & $< 5 \times 10^6$  & $ 3 \times 10^7$ & \\
 \hline
	\hline
	\multirow{3}*{Energy Budget [erg]} & \multicolumn{1}{c}{Thermal} & \multicolumn{2}{c ||}{Bulk Kinetic} & \multicolumn{2}{c}{Thermal} & \multicolumn{3}{c}{Bulk Kinetic\footnotemark[5]} \\
& \multicolumn{1}{c}{(Plasma in halo)} & \multicolumn{2}{c ||}{(Shocked H$\,${\sc i} gas)} & \multicolumn{2}{c}{(X-ray emitting plasma)} & \multicolumn{3}{c}{(Turbulent $\rm H_2$)} \\
& $ 1.7 \times 10^{55}$ & \multicolumn{2}{c ||}{$ 1 - 3 \times 10^{56}$} &  \multicolumn{2}{c}{$4 \times 10^{55}$} & \multicolumn{3}{c}{$\gtrsim 1.2 \times 10^{56}$} \\
	\hline
	\hline
  \multirow{2}*{Flux [W m$^{-2}$]} & \multicolumn{1}{c}{X-rays} & & & \multicolumn{1}{c}{X-rays} & \multicolumn{1}{c}{H$\, \alpha$} &   \multicolumn{1}{c}{O{\sc i}} & \multicolumn{1}{c}{H$_{2}$ \footnotemark[4]}\\
& $2.3 \times 10^{-18}$ & & & $1.1 \times 10^{-17}$ & $5.1 \times 10^{-18}$ & $3.6 \times 10^{-18}$ & $5.5 \times 10^{-17}$ \\
\hline
 \multirow{2}*{Luminosity\footnotemark[6]  [erg s$^{-1}$]} &  \multirow{2}*{$2.5 \times 10^{39}$} &  \multicolumn{2}{c||}{} &  \multirow{2}*{$1.2 \times 10^{40}$} &  \multirow{2}*{$5.4 \times 10^{39}$} &  \multirow{2}*{$3.8 \times 10^{39}$} &  \multirow{2}*{$5.8 \times 10^{40}$} \\
& & & & & & & &  \\
  \hline
	\hline
    \end{tabular}
   \caption[]{Mass, energy and luminosity budgets in front and behind the Stephan's Quintet galaxy-wide shock, directly inferred from observations. All the numbers are scaled to our aperture ($\mathcal{A} = 5.2 \times 2.1$~kpc$^{2}$) where \textit{Spitzer} observations were performed. The preshock gas is mainly the H{\sc i} gas contained in the tidal tail of NGC~7319. After the shock, the mass is distributed between the hot X-ray emitting plasma and the  $\rm H_2$ gas (see text for details). Before the shock, most of the energy available is the kinetic energy of the H{\sc i} gas that will be shocked. The shock splits the energy budget in two parts: thermal and kinetic energy. The former is stored in the hot plasma whereas the latter goes in turbulent motions that heat the $\rm H_2$ gas. Observations show that mechanical energy is the dominant energy reservoir.}
    \label{table_budgets}
\footnotetext[1]{XMM-Newton observations of the extended X-ray emission in the shock and the tidal tail \citep{2005A&A...444..697T}.}
\footnotetext[2]{H$\, \alpha$ and O{\sc i} optical line observations by  \citet{2003ApJ...595..665X}.}
\footnotetext[3]{Based on extrapolation of H{\sc i} observations in the tidal tail by \citet{2001AJ....122.2993S, Williams2002}.}
\footnotetext[4]{From Spitzer IRS observations. The $\rm H_2$ line flux is summed over the S(0) to S(5) lines \citep{2006ApJ...639L..51A}.}
\footnotetext[5]{This bulk kinetic energy is a lower limit because an unknown mass of turbulent cold molecular gas can contribute to this energy.}
\footnotetext[6]{Luminosities are indicated assuming a distance to SQ of 94~Mpc.}
\end{minipage}
\end{center}
\end{table*}

Observations show that the preshock and postshock gas are multiphase. We combine $\rm H_2$, H{\sc i}, H{\sc ii} and X-ray gas luminosities to estimate gas masses.

\subsubsection{Preshock gas}

The H{\sc i} preshock gas is contained on both the SQ tidal tail side (at a velocity of $\sim 6\,700$~km~s$^{-1}$) and the intruder galaxy side ($\sim 5\,700$~km~s$^{-1}$). On both sides, this gas is seen outside the shock area \citep{2001AJ....122.2993S, Williams2002}. On the SQ intra-group side, the tidal tail H{\sc i} column densities are in the range $N_{\rm H} = 1 - 3 \times 10^{20}$~cm$^{2}$. These two values bracket the range of possible preshock column densities. Neutral hydrogen observations of SQ show H{\sc i} at the intruder velocity to the South West (SW) of the shock area. The H{\sc i} column density of the SW feature and that of the tidal tail are comparable.
By multiplying the column densities by the area $\mathcal{A}$, we derive a total mass of H{\sc i} preshock gas in the aperture $\mathcal{A}$ of $2 - 6 \times 10^{7} \, \rm M_{\odot}$. 

Deep \textit{Chandra} and \textit{XMM-Newton} observations show a diffuse ``halo'' of soft X-ray emission from an extended region around the central shock region \citep{2005A&A...444..697T, O'Sullivan2008}. We consider that this emission is tracing the preshock plasma, which may have been created by a previous collision with  NGC~7320c \citep{2001AJ....122.2993S, 2003A&A...401..173T}.
The mass of hot gas, within the equivalent volume $V_{\rm \mathcal{A}}$ associated with the aperture $\mathcal{A}$, is derived from the X-ray luminosity and the temperature $T_{\rm X}$:
\begin{equation}\label{Eq_massX}
M_{\rm X} \simeq 6 \times 10^{6} \left( \frac{L_{\rm X}}{2.5 \times 10^{39}} \right)^{1/2} \left( \frac{3 \times 10^{-23}}{\Lambda (T_{\rm X})}\right)  ^{1/2} \left( \frac{l_{z}}{2  \, \rm kpc} \right)^{1/2} \ \rm M_{\odot},
\end{equation}
where $L_{\rm X}\, \rm [erg\,s^{-1}]$ is the X-ray luminosity, $\Lambda (T_{\rm X}) \, \rm [erg\,s^{-1}\,cm^{3}]$ the gas cooling efficiency at the plasma temperature $T_{\rm X}$. 
The X-ray luminosity in the halo is derived from  \citet{2005A&A...444..697T} and is  $2.5 \times 10^{39}$erg$\,s^{-1}$. We assume a solar metallicity \citep{2003ApJ...595..665X} and a cooling efficiency of $\Lambda (T_{\rm X}) = 3 \times 10^{-23} \,$erg~s$^{-1}$~cm$^{3}$ at $T_{\rm X} = 5 \times 10^6$~K \citep[see Appendix~\ref{subsec_coolfunct_plasma} and][for more details]{2007ApJS..168..213G}.

\subsubsection{Postshock gas}

The galaxy-wide collision creates a multi-phase interstellar medium where molecular and H{\sc ii} gas are embedded in a hot X-ray emitting plasma. 

No postshock H{\sc i} gas is detected in the shocked region where $\rm H_2$ was observed. We derive an upper limit of $ 5 \times 10^6\,$M$_{\odot}$ based on the lowest H{\sc i} contour close to the shock. A two-temperature fit of the $\rm H_2$ excitation diagram gives $T=185$~K and $T=675$~K for the two gas components. Most of the gas mass is contained in the lower temperature component, which represents a mass of warm $\rm H_2$ of $3 \times 10^7\,$M$_{\odot}$  \citep{2006ApJ...639L..51A} within the aperture $\mathcal{A}$. Published CO observations do not constrain the total $\rm H_2$ mass in the preshock and postshock gas.
Given the uncertainties on the preshock gas masses, one cannot infer a precise mass of cold molecular gas. 

The mass of hot plasma is derived from the \citet{O'Sullivan2008} \textit{Chandra} observations in the region observed by \textit{Spitzer}. The absorption-corrected X-ray luminosity is $1.2 \times 10^{40}$~erg~s$^{-1}$ within the aperture $\mathcal{A}$. Using equation~\ref{Eq_massX}, this corresponds to a mass of $1.4 \times 10^7\,$M$_{\odot}$. This is $2 - 3$ times the mass of the preshock hot plasma. \citet{O'Sullivan2008} report variations of   the average proton density in the range $n_{\rm H} = 1.2 - 2.3 \times 10^{-2}  \,$cm$^{-3}$. The density peaks close to the area $\mathcal{A}$. We adopt a hot plasma density of $n_{\rm H}  \simeq 0.02 \,$cm$^{-3}$ where $\rm H_2$ has been detected. This is in agreement with the value obtained by \citet{2005A&A...444..697T}.
The postshock gas pressure is given by
\begin{equation}
\frac{P_{\rm ps}}{k_{\rm B}} = 2.2  \times 10^5 \, \frac{0.6 \times m_{\rm H}}{\mu} \left( \frac{n_{\rm H}}{0.02 \, \rm cm^{-3}}\right) \left( \frac{T_{\rm ps}}{5 \times 10^{6} \, \rm K} \right)  \ \ \rm K~cm^{-3}.
\end{equation}
In the tenuous gas this pressure remains almost constant during the collision time scale since the plasma does not have the time to cool nor to expand significantly.
The isobaric gas cooling time scale can be estimated as the ratio of the internal energy of the gas to the cooling rate:
\begin{equation}\label{eq_gascooling_timescale}
t_{\rm cool} = \frac{5/2 \, k_{\rm B} \, T_{\rm X}}{n_{\rm H} \, \Lambda (T_{\rm X})} \simeq 9 \times 10^{7} \, \left( \frac{T_{\rm X}}{5 \times 10^{6} \, \rm K} \right)^{1.54} \, \left( \frac{0.02\, \rm cm^{-3}}{n_{\rm H}} \right)\  \rm yr.
\end{equation}
This cooling time is significantly longer than the age of the shock. 
In the right hand side of Eq.~\ref{eq_gascooling_timescale}, we use the power-law fit of the $Z=1$ cooling efficiency given in \citet{2007ApJS..168..213G} \citep[consistent with][]{1993ApJS...88..253S}. 
If the metallicity of the pre-shocked tidal filament was sub-solar, this would increase the cooling time of the gas still further. \citet{O'Sullivan2008} found a best fit to their X-ray spectrum for $Z=0.3 \, Z_{\odot}$.

The mass of the ionized H{\sc ii} gas is derived from the emission measure of H$\, \alpha$ observations:
\begin{equation}
M_{\rm H\begin{scriptsize}II\end{scriptsize}} \simeq 1.2 \times 10^{6} \left(  \frac{\mathcal{F}_{\rm H\alpha}}{5.1 \times 10^{-18}} \right)  \left( \frac{T}{10^{4}~\rm K}\right)  ^{0.92} \left(  \frac{10 \rm \, cm^{-3}}{n_{\rm H}} \right)  \, \rm M_{\odot}
\end{equation}
where $\mathcal{F}_{\rm H\alpha}$~[W~m$^{-2}$] is the flux of the
$H_{\alpha}$ line scaled to the aperture $\mathcal{A}$ and $n_{\rm H}$
the gas proton density. For SQ, the observed $H_{\alpha}$ flux is
$\mathcal{F}_{\rm H\alpha} = 5.1 \times 10^{-18}\,$W~m$^{-2}$
\citep{2003ApJ...595..665X}.  At the SQ postshock pressure, the gas
density at $10^4$~K is $n_{\rm H} \simeq 10$~cm$^{-3}$. 

\subsection{Extinction and UV field}\label{subsec_UVfield}

\citet{2003ApJ...595..665X} have interpreted their
non-detection of the H$_{\beta}$ line as an indication of significant
extinction ($E(\rm H_{\beta} - H_{\alpha}) > 1.2$, i.e. $A(H_{\alpha})
> 2.4$). 
However, more sensitive spectroscopy\footnote{observations made at the Cerro Tololo Inter-American Observatory (CTIO)} by P.-A. Duc et al. (private communication) show that the  H$_{\beta}$ line is detected over the region observed with \textit{IRS} at two velocity components ($\sim 5\,700$ and $6\,300$~km~s$^{-1}$, corresponding to the intruder and the intra-group gas velocities). 
These observations show that the extinction is much higher ($A_V = 2.5$) for the low-velocity gas component than for the high-velocity one. There are also large spatial variations of the extinction value for the intruder gas velocity close to this position, suggesting clumpiness of the gas within the disturbed spiral arm of NGC~7318b.
An extinction of $A_V=0.3-0.9$ is derived from the Balmer decrement for the intra-group tidal tail velocity, with much lower spatial variations. After subtraction of the foreground galactic extinction ( $A_{V} = 0.24$), this gives a SQ ridge extinction $A_{V}$ in the range $0.1 - 0.7$.


A low extinction for the H$_2$-rich, intra-group gas is also consistent with measured UV and far-IR dust surface brightness observations.  Using the published
value of FUV extinction \citep{2005ApJ...619L..95X} and the
\citet{Weingartner2001} extinction curve, we derive $A_V = 0.29$, which is in
agreement with the new Balmer-decrement study.  \textit{GALEX}
observations show that the radiation field is $G \sim 1 - 2 \, G_0$ in
the shocked region, where $G_0$ is the interstellar radiation field
(ISRF) in Habing units\footnote{$G_0 = 2.3 \times 10^{-3}$~erg~s$^{-1}$~cm$^{-2}$ at $1\,530$~\AA, \citet{Habing1968}}. We obtain the higher value using the flux listed in Table~1 of \citet{2005ApJ...619L..95X}. This flux is measured over an area much larger than our aperture $\mathcal{A}$. It
provides an upper limit on the UV field in the shock because this area
includes star forming regions in the spiral arm of NGC 7318b. ISO
observations show a diffuse 100$\, \mu$m emission at 2.2~MJy~sr$^{-1}$
level in the shocked region \citep{2003ApJ...595..665X}. Using the Galactic 
model of \citet{Draine2007} for $G = 2 \, G_0$, the 100$\, \mu$m FIR
brightness corresponds to a gas column density of $1.5 \times
10^{20}$~cm$^{-2}$. For $G = G_0$, we obtain $N_{\rm H} = 3.3 \times
10^{20}$~cm$^{-2}$, which corresponds to $A_V = 0.18$ for the Solar
neighbourhood value of the $A_V / N_{\rm H}$ ratio. Since this value is low, no extinction correction is applied.


\subsection{Energy reservoirs}\label{subsec_energy_reservoirs}

The bulk kinetic energy of the pre- and postshock gas are the dominant terms of the energy budget of the SQ collision (Table~\ref{table_budgets}). Assuming that the preshock H{\sc i} mass is moving at $V_{\rm s} = 600$~km~s$^{-1}$ in the center of mass rest frame, the preshock gas bulk kinetic energy  is  $1 - 3 \times 10^{56}$~erg. Note that this is a lower limit since a possible contribution from preshock molecular gas is not included. The thermal energy of the X-ray plasma represents $\sim 10\,$\% of the preshock kinetic energy.

The shock distributes the preshock bulk kinetic energy into postshock thermal energy and bulk kinetic energy.
The \emph{Spitzer} IRS observations of SQ  \citep{2006ApJ...639L..51A} show that the 0-0~S(1) 17~$\mu$m $\rm H_2$ line is resolved with a width of $870 \pm 60 \,$km~s$^{-1}$ (FWHM), comparable to the collision velocity.
We assume that the broad resolved line represents $\rm H_2$ emission with a wide range of velocities, by analogy with the O{\sc i}~[$6\,300$~\AA] emission seen in the same region by \citet{2003ApJ...595..665X} (see \citet{2006ApJ...639L..51A} for discussion). Then the results imply a substantial kinetic energy carried by the postshock gas. The kinetic energy of the $\rm H_2$ gas is estimated by combining the line width with the warm $\rm H_2$ mass. This is a lower limit since there may be a significant mass of $\rm H_2$ too cold to be seen in emission.
The $\rm H_2$ kinetic energy is  more than 3 times the amount of thermal energy of the postshock hot plasma. This implies that most of the collision energy is not dissipated in the hot plasma, but a substantial amount is carried by the warm molecular gas.

Radio synchrotron observations indicate that the postshock medium is
magnetized. Assuming the equipartition of energy between cosmic rays
and magnetic energy, the mean magnetic field strength in the SQ
shocked region is $\simeq 10\, \mu$G \citep{2003ApJ...595..665X}.  The
magnetic energy and thereby the cosmic ray energy contained in the
volume $V_{\mathcal{A}}$ is $\frac{B^2}{8\, \pi} \, V_{\mathcal{A}} =
2.7 \times 10^{54}$~erg. This energy is much smaller than the bulk
kinetic energy of the $\rm H_2$ gas. 

To conclude, the dominant postshock energy reservoir is the
  bulk kinetic energy of the molecular gas. We thus propose that the
  observed $\rm H_2$ emission is powered by the dissipation of this
  kinetic energy. The ratio between the kinetic energy of the
warm H$_2$ gas in the center of mass frame and the $\rm H_2$
luminosity is:
\begin{equation}
t_{\rm diss} \simeq 10^8  \left( \frac{1.5 \times 10^7 ~ \rm L_{\odot}}{L_{\rm H_2}} \right) \, \rm yr.
\end{equation}
The  $\rm H_2$ luminosity of $1.5 \times 10^7 ~ \rm L_{\odot}$ is summed over the S(0)-S(5) lines. This dissipation timescale is of the order of the cooling time of the hot X-ray emitting plasma . 
Therefore, the dissipation of the kinetic energy of the SQ collision can power the $\rm H_2$ emission on a timescale more than one order of magnitude longer than the collision age ($t_{\rm coll} \sim 5 \times 10^{6} \,$yr). The timescale of energy dissipation will be larger if the total amount of molecular gas is larger than the warm  $\rm H_2$   mass derived from \textit{Spitzer} observations. On the other hand, it will be smaller by a factor 2 if we consider cooling through the emission of the H{\sc ii} gas.

The line emission of the ionized gas could also contribute to the dissipation of energy.
We estimate the total luminosity of the postshock H{\sc ii} gas from the H$\, \alpha$ luminosity \citep{Heckman1989}:
\begin{equation}
L_{\rm H{\tiny \, II}} = \frac{13.6 \rm ~ eV}{0.45 \times E_{\rm H\alpha}} \, L_{\rm H\alpha} \ ,
\end{equation}
where $E_{\rm H\alpha}$ is the energy of an $H\alpha$ photon. Therefore, the luminosity of the postshock H{\sc ii} gas is $L_{\rm H{\tiny \, II}} = 2.1 \times 10^7$~L$_{\odot}$ within the aperture $\mathcal{A}$. This represents $16 \times L_{\rm H\alpha}$ and is comparable to the $\rm H_2$ luminosity.

\begin{figure*}
\begin{center}
 \includegraphics[width = \textwidth]{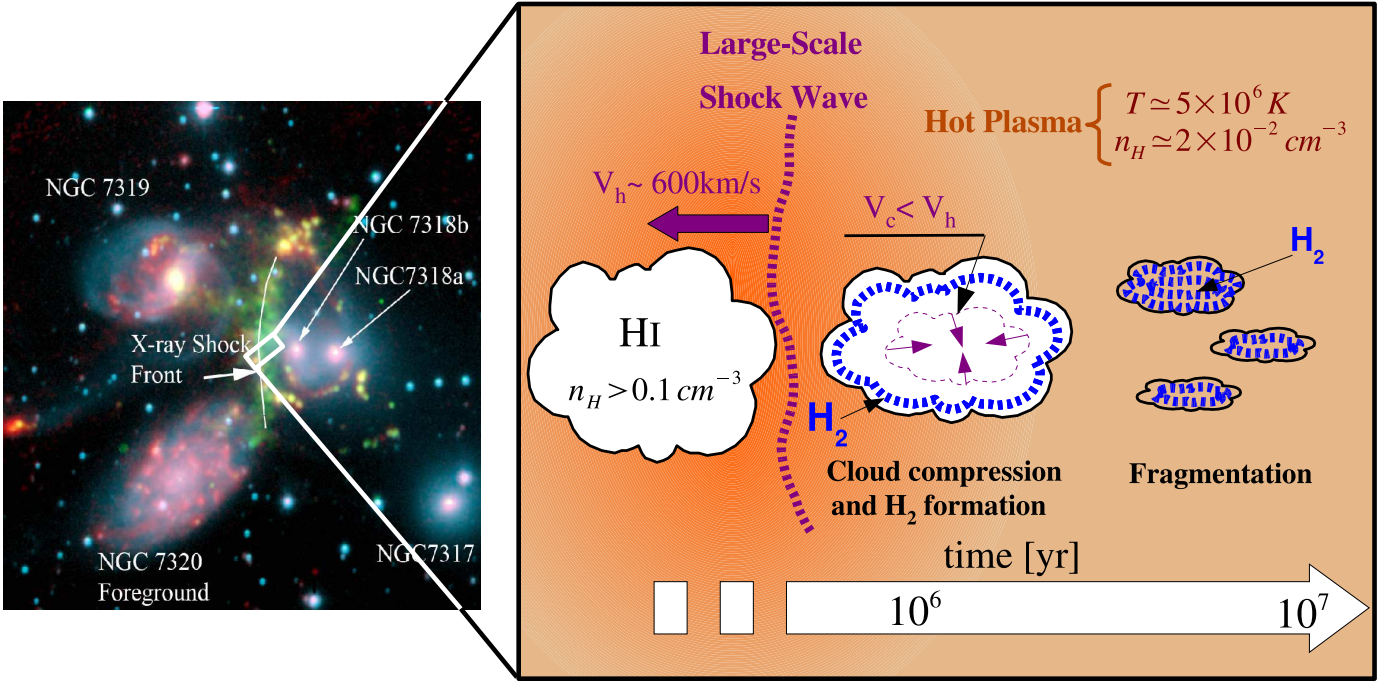}
 \caption{A schematic picture of a galactic wide shock in an inhomogeneous medium. \emph{Left:} A three data sets image of the Stephan's Quintet system (visible red light (blue), H$\alpha$ (green) from the Calar Alto Observatory in Spain, and 8-micron infrared light (red) from Spitzer's InfraRed Array Camera (IRAC). The white box represents the aperture $\mathcal{A}$ where \textit{Spitzer} spectroscopic observations were performed. \emph{Right:} Schematic view of the proposed scenario for the $\rm H_2$ formation in the SQ galaxy-wide collision. In the hot/tenuous medium, the shock is propagating fast ($V_h \sim 600$~km~s$^{-1}$), whereas in the  clouds a lower velocity shock ($V_c < V_h$) is driven. The cloud compression triggers gas cooling and $\rm H_2$ formation in the postshock clouds. Thermal instability and dynamical interaction with the hot plasma lead to molecular gas fragmentation (see \S~\ref{subsec_clouds_evolution}).}
\label{Fig_picture_shocks}
 \end{center}
  \end{figure*}

\section{Why is H$\bf _2$ present in the postshock gas?}
\label{section_scenario}

Observations show that the pre-collision medium is \textit{inhomogeneous}. The gas in the SQ group halo is observed to be structured with H{\sc i} clouds and possibly $\rm H_2$ clouds embedded in a teneous, hot, intercloud plasma (section~\ref{subsec_context}). On the intruder side, the galaxy interstellar medium is also expected to be multiphase.
We consider a large-scale collision between two flows of multiphase dusty gas (Fig.~\ref{Fig_picture_shocks}). 
In this section we quantify $\rm H_2$ formation within this context. The details of the microphysics calculations of the dust evolution, gas cooling and $\rm H_2$ formation are given in Appendices~\ref{appendix_dust_evol}, \ref{appendix_hotgascooling} and \ref{appendix_H2}.

\subsection{Collision between two dusty multiphase gas flows}\label{subsec_gal_shocks}


This section introduces a schematic description of the collision that allows us to quantify the $\rm H_2$ formation timescale.
The collision drives two high-velocity large-scale shocks that propagate through the teneous volume filling plasma, both into the tidal arm and into the intruder. 
The rise in the intercloud pressure drives slower shocks into the clouds \citep[e.g.][]{1975ApJ...195..715M}. 
The velocity of the transmitted shock into the clouds, $V_{\rm c}$,  is of order 
$\displaystyle V_{\rm c} \simeq \sqrt{n_{\rm h} / n _{\rm c}} \,  V_{\rm h} \ ,$
where $V_{\rm h}$ is the shock velocity in the hot intercloud medium, and $n_{\rm c}$ and $n _{\rm h}$ are the densities of the cloud and intercloud medium\footnote{More detailed expressions are given in \cite{Klein1994}}, respectively.
Each cloud density corresponds to a shock velocity. 

The galaxy collision generates a range of shock velocities and postshock gas temperatures. The state of the postshock gas is related to the preshock gas density. Schematically, low density ($n_{\rm H} \lesssim 10^{-2}\,$cm$^{-3}$) gas is shocked at high speed ($\gtrsim~600\,$km~s$^{-1}$) and accounts for the X-ray emission. The postshock plasma did not have time to cool down significantly and form molecular gas since the collision was initiated.
H{\sc i} gas ($n_{\rm H}  \gtrsim 3 \times 10^{-2}$) is heated to lower ($ \lesssim10^6$~K) temperatures. This gas has time to cool. Since $\rm H_2$  forms on dust grains, dust survival is an essential element of our scenario (see Appendix~\ref{appendix_dust_evol}).
The gas pressure is too high for warm neutral H{\sc i}  to be in thermal equilibrium. Therefore, the clouds cool to the temperatures of the cold neutral medium  and become molecular.

The time dependence of both the gas temperature and the dust-to-gas mass ratio, starting from gas at an initial  postshock temperature $T_{\rm ps}$, has been calculated.
We assume a galactic dust-to-gas ratio, a solar metallicity and equilibrium ionization. In the calculations, the thermal gas pressure is constant. As the gas cools, it condenses. 
Below $\sim 10^4\, $K, the H{\sc ii} gas recombines, cools through optical line emission, and becomes molecular. The gas and chemistry network code (Appendix~\ref{appendix_H2}) follows the time evolution of the gas that recombines. The effect of the magnetic field on the compression of the gas has been neglected in the calculations and we assume a constant thermal gas pressure.

Under the usual flux-freezing ideas, the magnetic field strength should increase with
volume density and limit the compression factor. However, observations show that the magnetic field in the cold neutral medium (CNM) is comparable to that measured  in lower density, warm ISM components  \citep{Heiles2005}. 3D MHD numerical simulations of the supernovae-driven turbulence in the ISM also show that gas compression can occur over many orders of magnitude, without increasing the magnetic pressure \citep[e.g.][]{Mac2005, deAvillez2005}. 
The reason is that the lower density ($n_{\rm H} \lesssim 10^3$~cm$^{-3}$) gas is magnetically sub-critical and mainly flows along the magnetic field lines without compressing the magnetic flux. 
The magnetic field intensity increases within gravitationally bound structures at higher gas densities ($n_{\rm H} \gtrsim 10^3$~cm$^{-3}$, \citep[e.g.][]{Hennebelle2008}. This result is in agreement with observations of the magnetic field intensity in molecular clouds by \citet{Crutcher1999} and \citet{Crutcher2003}.



\subsection{Gas cooling, dust destruction and $\rm H_2$ formation timescales}\label{subsec_timescales}

\begin{figure}[t]
   \includegraphics[angle=90, width = \columnwidth]{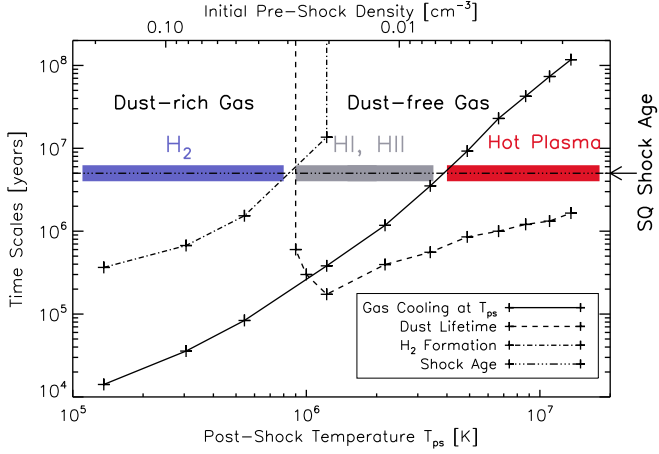}
   \caption{The multiphasic postshock medium. The dust lifetime, $\rm H_2$ formation and gas cooling (solid line) time scales in the SQ postshock gas are plotted as a function of the postshock temperature and preshock density (top axis). 
The state of the multiphase postshock gas is set by the relative values of these timescales at the SQ collision age ($\sim 5 \times 10^6\,$yr, indicated with the horizontal dashed line and marked with an arrow). By comparing these timescales one can define three gas phases,
marked by blue, grey and red thick lines. In particular, the comparison of the $\rm H_2$ formation time scale with the SQ collision age show where $\rm H_2$ molecules can form (blue thick line). }
   \label{Fig_timescales}%
\end{figure}

The physical state of the post-shock gas depends on the shock age, the gas
cooling, the dust destruction, and the $\rm H_2$ formation timescales. 
These relevant timescales are plotted as a function of the postshock
temperature in Fig.~\ref{Fig_timescales}.  
Given a shock velocity $V_{\rm h}$ in the low-density plasma, each value of the postshock temperature $T_{\rm ps}$ corresponds to a cloud preshock density $ n_{\rm c, p}$:
\begin{equation}
T_{\rm ps} = 2.4 \times 10^{5} \left(  \frac{n_{\rm h}}{0.02 \ \rm cm^{-3}}\right)  \left( \frac{0.1 \ \rm cm^{-3}}{ n_{\rm c, p}} \right) \left( \frac{V_{\rm h}}{600 \ \rm km~s^{-1}} \right) ^{2} \ \rm K.
\end{equation}

Dust destruction is included in the calculation of the H$_2$ formation
timescale.  The H$_2$ formation timescale is defined as the time when the H$_2$ fractional abundance reaches  90\% of its final (end of cooling) value, including the gas
cooling time from the postshock temperature. The dust
destruction timescale is defined as the time when 90\% of the dust
mass is destroyed and returned to the gas.  At $T < 10^{6}$~K, this
timescale rises steeply because the dust sputtering rate drops (see
Appendix~\ref{appendix_dust_evol} for details).  The dust destruction
timescale increases towards higher temperatures because, for a fixed
pressure, the density decreases.

The present state of the multiphase postshock gas is set by the
relative values of the three timescales at the SQ collision age,
indicated with the horizontal dashed line and marked with an arrow.
The ordering of the timescales depends on the preshock density and postshock gas temperature. By comparing these timescales one can define three gas phases, marked by blue, grey and red thick lines in Fig.~\ref{Fig_timescales}.
\begin{enumerate}
\item
\textit{The molecular phase~:} For $T_{\rm ps} \lesssim 8 \times 10^{5}$~K, the dust destruction timescale is larger than the collision age and the $\rm H_2$ formation timescale is lower.
This gas keeps a significant fraction of its initial dust content and becomes $\rm H_2$ (blue thick line).
\item  \textit{Atomic {\rm H{\sc i}} and ionized {\rm H{\sc ii}} gas phase~:} For intermediate temperatures ($8 \times 10^{5} < T_{\rm ps} < 3 \times 10^{6}$~K), the gas has time to cool down but looses its dust content.
The $\rm H_2$ formation timescale is greater than the collision age. This phase is indicated with the grey thick line.
\item \textit{X-ray emitting plasma~:} At high temperatures ($T_{\rm ps} > 3 \times 10^{6}$~K), the postshock gas is dust-free and hot. This is the X-ray emitting plasma indicated by the red thick line.
\end{enumerate}


In Fig.~\ref{Fig_timescales} it is assumed that, during the gas cooling, the molecular gas is in pressure equilibrium with the hot, volume filling gas. The gas pressure is set to the measured average thermal pressure of the hot plasma (HIM). 
Theoretical studies and numerical simulations show that the probability distribution function of the density in a supersonically turbulent isothermal gas is lognormal, with a dispersion which increases with the Mach number \citep{Vazquez-Semadeni1994, Padoan1997, Passot1998, Mac2005}. 
Based on these studies, the SQ postshock pressure in the turbulent cloud phase is likely to be widely distributed around the postshock pressure in the HIM ($2 \times 10^5$~K~cm$^{-3}$). The H$_2$ formation timescale would be larger in the low pressure regions and shorter in high pressure regions. 

\begin{figure}[t]
   \includegraphics[angle=90, width = \columnwidth]{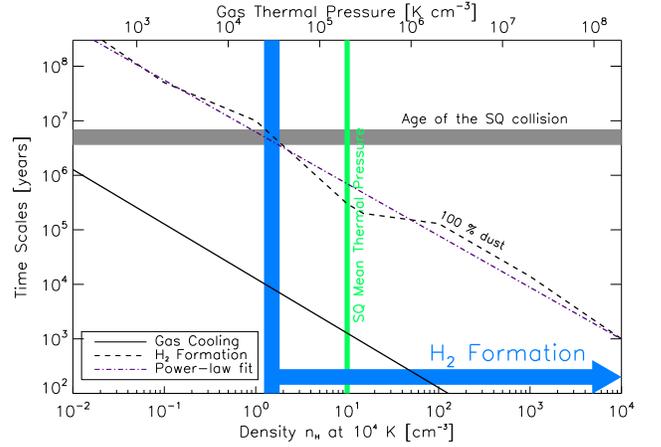}
   \caption{Cooling and H$_2$ formation timescales for ionized gas at $10^4$~K as a function of the gas density and pressure (top axis). 
The grey horizontal bar indicates the SQ collision age, and the blue vertical bar shows the gas thermal pressure at which the $\rm H_2$ formation timescale equals the collision age.
The vertical green bar indicates the  thermal pressure of the hot plasma. A power-law fit to the H$_2$ formation timescale is indicated (Eq.~\ref{eq_H2formationtime}). The H$_2$ formation timescale is shorter than the collision age for pressures larger than $1/10$ of the hot plasma pressure.}
   \label{Fig_timescales_1E4K_vs_nH_SQ}%
\end{figure}

To quantify this statement, we have run a grid of isobaric cooling models for different gas thermal pressures. Fig.~\ref{Fig_timescales_1E4K_vs_nH_SQ} shows the gas cooling time and the $\rm H_2$ formation timescale as a function of the gas thermal pressure.
A Solar Neighborhood value of the dust-to-gas ratio is assumed. 
The $\rm H_2$ formation time scale roughly scales as the inverse of the gas density at $10^{4}$~K and can be fitted by the following power-law:
 \begin{equation}\label{eq_H2formationtime}
t_{H_2} ~ \rm [yr]  \simeq 7 \times 10^{5} \, f_{\rm dust} \, \left( \frac{2 \times 10^{5} ~\rm [K\, cm^{-3}]}{P_{\rm th}}\right)^{0.95} \ \ , 
\end{equation}
where $\displaystyle P_{\rm th}$ is the thermal gas pressure and  $f_{\rm dust}$ the dust mass fraction remaining in the gas (In Fig.~\ref{Fig_timescales_1E4K_vs_nH_SQ}, $f_{\rm dust}=1$).

 Fig.~\ref{Fig_timescales_1E4K_vs_nH_SQ} shows that for a wide range of gas thermal pressures ($ > 3 \times 10^4$~K~cm$^{-3}$, i.e. $P_{\rm th}(\rm CNM) > \frac{1}{10} P_{\rm th}(\rm HIM)$),  $\rm H_2$ can form within less than the collision age ($ 5 \times 10^6$~yrs). 
Numerical simulations of supernovae-driven turbulence in the ISM show that this condition applies to a dominant fraction of the gas mass \citep[see Fig.~10 of][]{Mac2005}. Numerical studies by \citet{2007ApJ...659.1317G} show that the  pressure variations induced by the turbulence globally reduces the time required to form large quantities of H$_2$ within the Galactic CNM. 

\subsection{Cloud crushing and fragmentation timescales}\label{subsec_clouds_evolution}

In \S~\ref{subsec_timescales}, the microphysics of the gas has been quantified, ignoring the fragmentation of clouds by thermal, Rayleigh-Taylor and Kelvin-Helmholtz instabilities. In this section, we introduce the dynamical timescales associated with the fragmentation of clouds and compare them with the $\rm H_2$ formation timescale. 

\subsubsection{Fragmentation}

When a cloud is overtaken by a strong shock wave, a flow of background gas establishes around it, and the cloud interacts with the low-density gas which is moving relative to it  (see Fig.~\ref{Fig_picture_shocks}).
The dynamical evolution of such a cloud has been described in many papers \citep[e.g.][]{Klein1994, Mac1994}. Numerical simulations show that this interaction triggers clouds fragmentation.
The shock compresses the cloud on a \textit{crushing time} $t_{\rm crush} = R_{\rm c} / V_{\rm c}$, where $V_{\rm c}$ is the shock velocity in the cloud and $R_{\rm c}$ its characteristic preshock size.
The shocked cloud is subject to both Rayleigh-Taylor and Kelvin-Helmholtz instabilities.

Gas cooling introduces an additional fragmentation mechanism because the cooling gas is thermally unstable for temperatures smaller than $\sim 10^6\,$~K. 
The thermal instability generates complex inhomogeneous structures, dense regions that cool and low density voids \citep{2003ApJ...591..238S, Audit2005}. This fragmentation occurs on the gas cooling timescale. 
In most numerical simulations, the thermal instability is ignored because simplifying assumptions  are made on the equation of state of the gas (adiabatic, isothermal or polytropic  evolution).

\begin{figure}[t]
   \includegraphics[angle=90, width = \columnwidth]{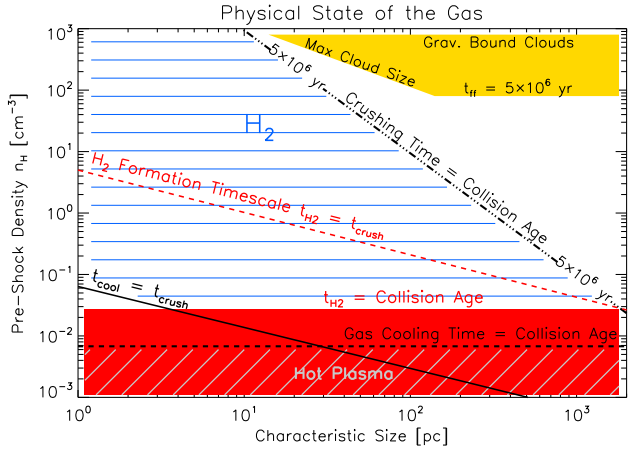}
\caption{The gas cooling and  $\rm H_2$ formation timescales are compared to the cloud crushing time and SQ collision age as a function of the cloud parameters (size and preshock density). The solid black line indicates where the gas cooling time $t_{\rm cool}$ equals the crushing time $t_{\rm crush}$. The black dashed line show where the cloud crushing timescale equals $5 \times 10^6$~yr (i.e. the SQ collision age). The red dashed line defines the location where the $\rm H_2$ formation timescale $t_{\rm H_2}$ equals the crushing timescale. The red area indicates where the $\rm H_2$ formation timescale is greater than the collision age, whereas the blue hatched zone show the parameter space where $\rm H_2$ can form. The yellow zone is the location of gravitationally bound, unstable clouds that have a free-fall time shorter than the collision age.}
   \label{Fig_cloud_survival}%
\end{figure}

In Fig.~\ref{Fig_cloud_survival}, the $\rm H_2$ formation and gas cooling timescales are compared to the crushing time and collision age, as a function of the initial density and size of the cloud. The pressure of the clouds is $2 \times 10^{5}$~K~cm$^{-3}$. 
The red zone indicates where the $\rm H_2$ formation timescale is larger than the SQ collision age. In this domain, clouds do not have time to form  $\rm H_2$. The blue hatched zone defines the  range of sizes and densities of the clouds which become molecular within $5 \times 10^{6}$~yr.
In this zone, the crushing time is larger than the cooling time. Gas fragments by thermal instability as the shock moves into the cloud.
$\rm H_2$ forms before the shock has crossed the cloud and therefore before Rayleigh-Taylor and Kelvin-Helmholtz instabilities start to develop.
The black dashed line is where the cloud crushing timescale equals the SQ collision age ($5 \times 10^6$~yr). To the right of this line, the clouds have not been completely overtaken and compressed by the transmitted shock.

Preshock Giant Molecular Clouds (hereafter GMCs) of
size $\gtrsim 30$~pc and mean density $\gtrsim 3 \times
10^2$~cm$^{-3}$, like those observed in spiral galaxies, 
fall within the yellow zone in Fig.~\ref{Fig_cloud_survival}. 
The bottom boundary corresponds to the density at which the free-fall time equals the age of the SQ collision. For densities higher than this limit, the free-fall time is shorter than the SQ collision age. The left boundary of this area is the maximum cloud size set by the Bonnor-Ebert criterion for a  pressure of $2 \times 10^{5}$~K~cm$^{-3}$. Clouds with larger sizes are gravitationally unstable. Such clouds are expected to form stars. 

A natural consequence of pre-existing GMCs would be
that the shock would trigger their rapid collapse and star formation
should rapidly follow \citep[e.g.][]{1992ApJ...387..152J}. The absence (or weakness) of these tracers in the region of the main shock \citep{2003ApJ...595..665X} suggests that fragments of preshock GMCs disrupted by star formation are not a major source of the observed postshock $\rm H_2$ gas. However, it is possible that some of the gas
in the intruder did contain pre-existing molecular material. The star formation regions seen to  the North and South of the ridge are consistent with this hypothesis. 

\subsubsection{Survival of the fragments}

The breakup of clouds can  lead to mixing of cloud gas with the hot background plasma  \citep{Nakamura2006} and thereby raises the question whether the cloud fragments survive long enough for H$_2$ to form. 
Fragments evaporate into the background gas when their size is such that
the heat conduction rate overtakes the cooling rate. The  survival of the fragments thus depends on their ability to cool \citep{Mellema2002, Fragile2004}.
Gas condensation and $\rm H_2$ formation are both expected to have  a stabilizing effect on the cloud fragments, because they increase the gas cooling power \citep{Le1999}.

So far, no simulations include the full range of densities and spatial scales and the detailed micro-physics which are involved  in the cloud fragmentation and eventual mixing with the background gas.
\citet{Slavin2006} reports an analytical estimate of the evaporation time scale of tiny H$\,${\sc i} clouds
embedded in warm and hot gas of several million years. This timescale is long enough for H$_2$ to form within the gas fragments lifetime.



\begin{figure*}
\begin{center}
  \includegraphics[angle=90, width = \textwidth]{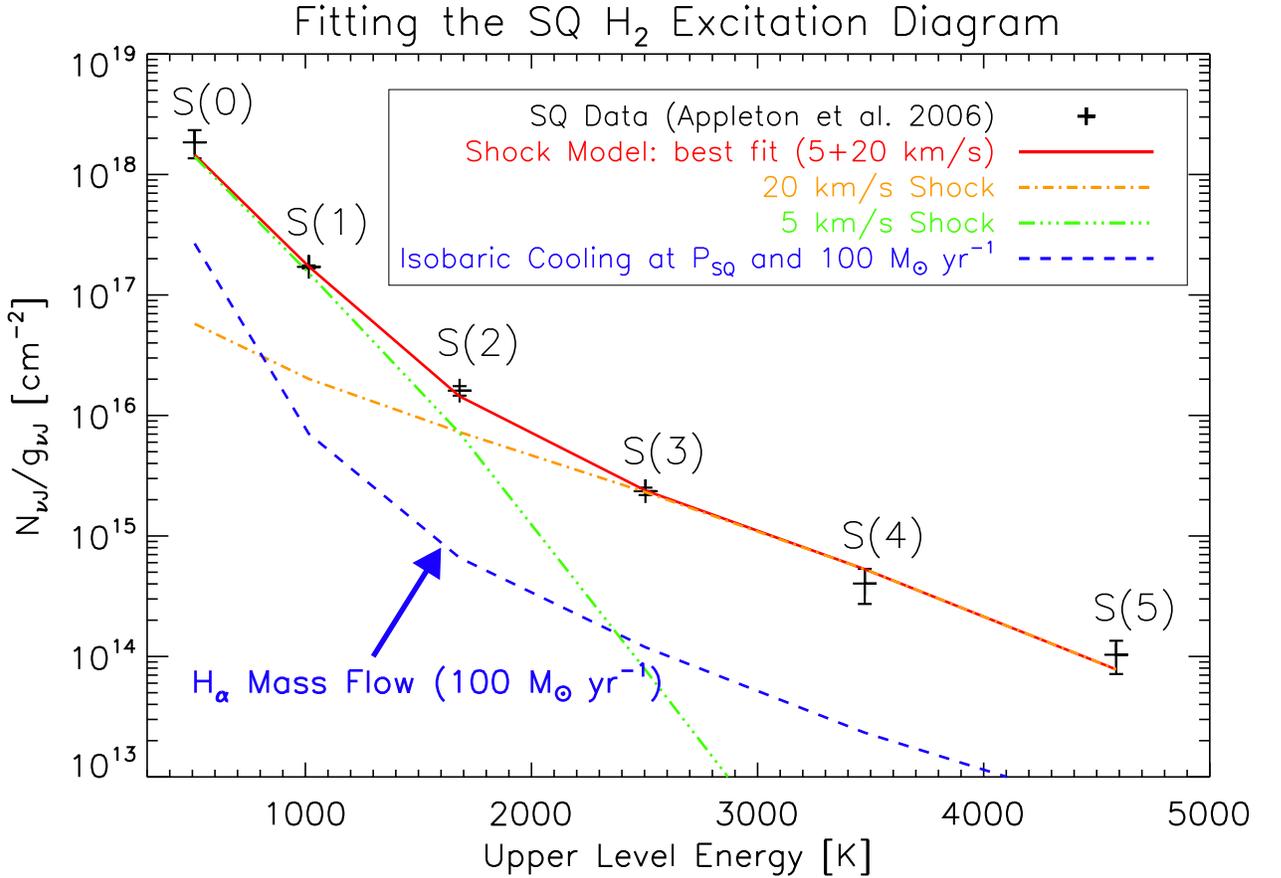}
   \caption{$\rm H_2$ excitation diagram fit by shock models. The data  \citep{2006ApJ...639L..51A} are the black crosses with their 1-$\sigma$ error bars.
The blue dashed line show the modeled excitation associated with the $ \rm H_2$ formation during the isobaric cooling of the postshock gas. The pressure is $2 \times 10^{5}$~K~cm$^{-3}$. The model column densities correspond to a mass flow of recombining gas of $100~\rm M_{\odot}~yr^{-1}$.
The green and orange dashed lines show  the contribution of $5$ and $20$~km~s$^{-1}$ shocks to the best fit model (red line), multiplied by the weights determined when fitting the line fluxes, respectively. Shocks are driven into molecular gas of preshock density $n_{\rm H} = 10^4\,$cm$^{-3}$. The red line shows the linear combination of the  two magnetic shocks (with velocities of $5$ and $20$~km~s$^{-1}$).}
    \label{Fig_H2excitDiag_SQ_isoP}
   \label{Fig_H2excitDiag_SQ}%
\end{center}
\end{figure*}

\section{ $\rm \bf H_2$ excitation}\label{sec_H2excitation}

This section addresses the question of $\rm H_2$ excitation. The models  (\S~\ref{subsec_MHDshocks}) and the results (\S~\ref{subsec_excit_diag_fitting_shocks}) are presented.

\subsection{Modeling the dissipation of turbulent kinetic energy by MHD shocks}
\label{subsec_MHDshocks}

Given that the dominant postshock energy reservoir is the bulk kinetic energy of the $\rm H_2$ gas (\S~\ref{subsec_energy_reservoirs}), the $ \rm H_2$ emission is assumed to be powered by the dissipation of turbulent kinetic energy into molecular gas. 
Radio synchrotron observations indicate that the postshock medium is
magnetized (\S~\ref{subsec_energy_reservoirs}).  In this context,  the dissipation of mechanical energy within  molecular gas is modeled by non-dissociative MHD shocks. 
In the absence of magnetic field, shocks are rapidly dissociative and much of the cooling of the shocked gas  occurs through atomic, rather than H$_2$, line emission. 
We quantify the required gas densities and shock velocities to account for the $\rm H_2$ excitation diagram.



  The $\rm H_2$ emission induced by low velocity shocks is calculated using an
updated version of the shock model of \citet{2003MNRAS.343..390F}.  In
these models, we assume a standard value for the cosmic ray ionization
rate of $\zeta = 5 \times 10^{-17}$~s$^{-1}$. In its initial state,
the gas is molecular and cold ($\sim 10$~K), with molecular abundances resulting from the output of our model for the isobaric cooling
(\S~\ref{subsec_cloudcoolingH2form} and
Fig.~\ref{Fig_species_time}). A grid of shock models has been computed for shock
velocities from $3 \ \rm to \ 40$~km~s$^{-1}$ with steps of $1$~km~s$^{-1}$, two values of the initial $\rm H_2$ ortho to para  ratio ($3$ and $10^{-2}$), and 3
different preshock densities ($n_{\rm H} = 10^2$, $10^3$, and
$10^4$~cm$^{-3}$).  We adopt the scaling of the initial magnetic strength with
the gas density, $B(\mu G) = b\, \sqrt{n_{\rm H}~[\rm cm^{-3}]}$
\citep{Crutcher1999, Hennebelle2008}. In our models, $b=1$. 

\subsection{Results: contribution of MHD shocks to the $\rm H_2$ excitation}\label{subsec_excit_diag_fitting_shocks}

Results are presented in terms of an H$_2$ excitation diagram, where the logarithm of the column densities $N_{\nu J} [\rm cm^{-2}]$, divided by the statistical weight $g_{\nu J}$, are plotted versus the energy of the corresponding level. The excitation diagrams are calculated for a gas temperature of 50~K, since the molecular gas colder than 50~K does not contribute to the H$_2$ emission.  Fig.~\ref{Fig_H2excitDiag_SQ_isoP} shows the $\rm H_2$ excitation diagram for both  observations and model. The SQ data from \citet{2006ApJ...639L..51A} is indicated with $1\sigma$ error bars. The $\rm H_2$ line fluxes have been converted to mean column densities within our area $\mathcal{A}$. 

The observed rotational diagram is curved, suggesting that the H$_2$-emitting gas is a mixture of  components at different temperatures. The data cannot be fitted with one single shock. At least two different shock velocities are needed.
A least-squares fit of the observed $\rm H_2$ line fluxes to a linear combination of 2 MHD shocks at different velocities $V_{s1}$ and $V_{s2}$ has been performed.
For a preshock density of $10^4$~cm$^{-3}$, the best fit is obtained for a combination of $5$ and $20$~km~s$^{-1}$ shocks with an initial value of the ortho to para ratio of $3$ (Fig.~\ref{Fig_H2excitDiag_SQ} and Table~\ref{table_results}). Note  that no satisfactory fit  could be obtained with the low value of the initial H$_2$ ortho-to-para ratio.
The contributions of the $5$ (green dashed line) and $20$~km~s$^{-1}$  (orange dashed line) shocks to the total column densities are also plotted.
Qualitatively, the $5$~km~s$^{-1}$ component dominates the contribution to the column densities of the upper levels of S(0) and S(1) lines.
The $20$~km~s$^{-1}$ component has a major contribution for the upper levels of S(3) to S(5). For S(2), the 2 contributions are comparable.
The red line shows the weighted sum of the two shock components (best fit). The warm H$_2$ pressure $\rho \, V^{2}$ in the two shocks is $4.5 \times 10^{7}$~K~cm$^{-3}$ (respectively $5.6 \times 10^{8}$~K~cm$^{-3}$) for the $5$~km~s$^{-1}$ (resp. $20$~km~s$^{-1}$) shock.

Our grid of models shows that this solution is not unique.
If one decreases the density to  $10^3$~cm$^{-3}$, we find that one can fit the observations with a combination of MHD shocks (at $9$ and $35$~km~s$^{-1}$). 
If one decreases the density to  $10^2$~cm$^{-3}$, the rotational $\rm H_2$ excitation cannot be fitted satisfactorily. At such low densities MHD shocks fail to reproduce the S(3) and S(5) lines because the critical densities for rotational H$_2$ excitation increases steeply with the J rotational quantum number \citep{Le1999}.
Both warm and dense (i.e. high pressure)  gas is needed to account for emission from the higher J levels.


The data fit gives an estimate of the shock velocities required to account for the H$_2$ excitation. The velocities are remarkably small with respect to the SQ collision velocity.
The fact that energy dissipation occurs over this low velocities range is an essential key to account for the importance of H$_2$ cooling. High velocity shocks would dissociate H$_2$ molecules.

\section{The cycling of gas across ISM phases}
\label{sec_cycle}

An evolutionary picture of the inhomogeneous postshock gas emerges from our interpretation of the data. The powerful $\rm H_2$ emission is part of a broader astrophysical context, including optical and X-ray emission which can be understood in terms of mass exchange between gas phases.   A schematic cartoon of the evolutionary picture of the postshock gas is presented in Fig.~\ref{Fig_mass_flows2}.
It illustrates the cycle of mass between the ISM phases that we detail here.
 This view of the SQ postshock gas introduces a physical framework that may apply to H$_2$ luminous galaxies in general.

\subsection{Mass flows}\label{subsec_MassFlows}

The $\rm H_2$ gas luminosity is proportional to the mass of gas that cools per unit time, the so-called mass flow. We compute the mass flow associated with one given shock dividing the postshock column density down to a gas temperature of 50~K by the cooling time needed to reach this temperature. Gas cooler than 50~K does not contribute to the $\rm H_2$ emission.

The two-components shock model has been used to deduce the mass flows needed to fit the data.
The cooling timescale (down to 50~K) is set by the lowest shock velocity component. For a shock velocity of 5~km~s$^{-1}$, the cooling
time is $\sim 2 \times 10^{4}$~yr.
This gives a  total mass flow required to fit the $\rm H_2$ emission of $\dot{M_3} = 5.7 \times 10^3\,$M$_{\odot}$~yr$^{-1}$.
For comparison, for the 20~km~s$^{-1}$ shock, the cooling time down to 50~K is $\sim 4 \times 10^{3}$~yr and the associated mass flow is $\dot{M} = 270\,$M$_{\odot}$~yr$^{-1}$.

These mass flows are compared to the mass flow of recombining gas that can be estimated from the H$\, \alpha$ luminosity  $L_{\rm H\alpha}$:
\begin{equation}
L_{\rm H\alpha} \, = \, 0.45 \times E_{\rm H\alpha} \times \Phi_{\rm Rec} \ ,
\end{equation}
where $E_{\rm H \alpha}$ is the energy of an H$\, \alpha$ photon and $\Phi_{\rm Rec}$ is the number of hydrogen atoms that recombine per second \citep{Heckman1989}.
For our reference area $\mathcal{A}$, we find a mass flow of recombining gas $\dot{M}_{\rm H _\alpha} = m_{\rm H} \, \Phi_{\rm Rec} \simeq 100~\rm M_{\odot}~yr^{-1}$.
If the ionizing rate of warm gas is smaller than  $\Phi_{\rm Rec}$, most of the recombining gas cools.
In this case $\Phi_{\rm Rec}$ provides the estimate of the mass flow needed to compute the cooling gas H$_2$ luminosity.
$\Phi_{\rm Rec}$ is an upper limit if we suppose that some of the H$_\alpha$ emission arises from collisional excitation of neutral hydrogen or if some of the warm gas is re-ionized before it has time to cool.

The H$_2$ emission integrated over the isobaric cooling sequence is calculated. 
The blue dashed curve of Fig.~\ref{Fig_H2excitDiag_SQ} corresponds to the model results for a mass flow of
$\dot{M} = 100~\rm M_{\odot}~yr^{-1}$ and a pressure $P_{\rm SQ} = 2\times  10^5$~K~cm$^{-3}$.
The model results do not depend much on the pressure. The model reproduces well the shape of the observed excitation diagram, but the emission is a factor $\sim 25$  below the observations. The model also fails to reproduce the [O{\sc i}] $6\,300$~\AA~line emission by a factor of 7.
Therefore, the cooling of the warm H{\sc i} gas produced by H{\sc ii} recombination cannot reproduce the observed $\rm H_2$ emission nor the  [O{\sc i}] $6\,300$~\AA~line emission.
A larger amount of energy needs to be dissipated within the molecular gas to account for the $\rm H_2$ emission. This is why the fit total mass flow associated with the 2-component MHD shocks  is much larger than the mass flow of recombining gas.

\begin{table*}
\begin{center}
\begin{minipage}[t]{\textwidth}
 \renewcommand{\footnoterule}{}
\def\thefootnote{\alph{footnote}}
\centering
    \begin{tabular}{ c | c c c c | c c c c c}
	\hline
	\hline
  & & & & & \multicolumn{5}{c}{{\sc Flux of Spectral Features} ($10^{-18}$ W~m$^{-2}$)} \\
\cline{6-10}
    & Mass Flow & $P/k_{\rm B}$ & $V_s$ \footnotemark[1] & Cooling Time\footnotemark[2] & \multirow{2}*{O{\sc i} [$63~\mu$m]} & \multirow{2}*{$\rm H_2$ rot.\footnotemark[3]} &  \multirow{2}*{$\rm H_2$ 1-0 S(1)} & \multirow{2}*{O{\sc i} $6\,300$~\AA \footnotemark[4] } & \multirow{2}*{$\rm H_{\alpha}$ \footnotemark[4]}  \\
 & [M$_{\odot}$~yr$^{-1}$]  & [K~cm$^{-3}$] & [km~s$^{-1}$] & [yr] & & & & \\
  \hline
Observations & & & & & & $55 \pm 3$ & & 3.6 & 5.3 \\
  \hline
\multirow{2}*{Isobaric Cooling} & $100$ (WIM, H$_\alpha$)\footnotemark[5]  & $2 \times 10^5$ & & $1.2 \times 10^4$& 3.9 & 1.4 & 0.05 & 0.5 & 5.3 \\
  & $640$ (WNM, O{\sc i})\footnotemark[6] & $2 \times 10^5$ & & $1.5 \times 10^5$  & 25.8 & 9.1 & 0.3 & 3.6 &  \\
\hline
\multirow{2}*{Shock Models} & $5.7 \times 10^3$ & $4.5 \times 10^7$ & 5  & $1.8 \times 10^4$ & 13 & 27 & 0.3 & $10^{-5}$& \\
& $270$ & $5.6 \times 10^8$ & 18  & $3.7 \times 10^3$ & 0.04 & 25 & 0.01 &  $5 \times 10^{-3}$& \\
  \hline
	\hline
    \end{tabular}
   \caption[]{Overview of the emission of the SQ postshock gas, from observations and model predictions. All line fluxes are scaled to the aperture $\mathcal{A} = 11.3 \times 4.7$~arcsec$^{2}$. For isobaric cooling calculations, the pressure is set to the SQ postshock gas value. For MHD shock models, both contributions of the 2 shock velocities to the emission are indicated, as well as the pressure for the warm $\rm H_2$ phase in the shock. These models are the same as those used in Fig.~\ref{Fig_H2excitDiag_SQ}}
    \label{table_results}
\footnotetext[1]{MHD Shock velocity.}
\footnotetext[2]{Cooling time computed down to 50~K.}
\footnotetext[3]{Sum of the $\rm H_2$ S(0) to S(5) rotational lines (from \citet{2006ApJ...639L..51A}).}
\footnotetext[4]{From optical observations by \citet{2003ApJ...595..665X}.}
\footnotetext[5]{Mass flow derived from H$\, \alpha$ observations, ignoring ionization. It represents the isobaric cooling of the recombining H{\sc ii} gas.}
\footnotetext[6]{Cooling mass flow derived from O{\sc i} observations and model calculations at the ambient SQ pressure. This mass flow does not depend much on the pressure since the critical density of the O{\sc i} line is high ($2 \times 10^6$~cm$^{-3}$).}
\end{minipage}
\end{center}
\end{table*}

\subsection{The mass cycle}\label{subsec_cycle}

\begin{figure}
\begin{center}
  \includegraphics[width = \columnwidth]{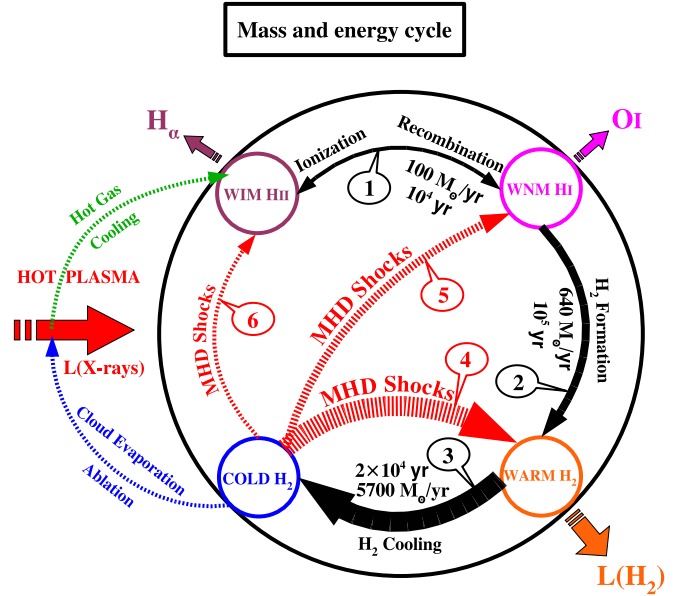}
   \caption{Schematic view at the evolutionary cycle of the gas proposed in our interpretation of optical and H$_2$ observations of Stephan's Quintet.
Arrows represent the mass flows between the H{\sc ii}, warm H{\sc i}, warm and cold H$_2$ gas components. They are numbered for clarity (see text). The dynamical interaction between gas phases drives the cycle. The values of the mass flows and associated timescales are derived from the $\rm H \alpha$, [O{\sc i}~$6\,300$~\AA], $\rm H_2$ observed luminosities and model calculations (cf. Table~\ref{table_results}).
Heating of the cold H$_2$ gas (red dashed arrows) is necessary to account for the increasing mass flow from the ionized gas to cold H$_2$  phases. }
   \label{Fig_mass_flows2}%
\end{center}
\end{figure}

This section describes the mass cycle of  Fig.~\ref{Fig_mass_flows2}. 
Black and red arrows represent the mass flows between the H{\sc ii}, warm H{\sc i}, warm and cold H$_2$ gas components of the postshock gas. 
The large red arrow to the left symbolizes the relative motion between the warm and cold gas and the surrounding plasma. Each of the black arrows is labeled with its main associated process: gas recombination and ionization (double arrow number 1), H$_2$ formation (2) and H$_2$ cooling (3). The values of the mass flows and the associated timescales are derived from observations and
our model calculations (\S~\ref{subsec_MassFlows} and Table~\ref{table_results}).


A continuous cycle through gas components is excluded by the increasing mass flow
needed to account for the $\rm H \alpha$, O{\sc i}, and $\rm H_2$ luminosities.
Heating of the cold H$_2$ gas towards warmer gas states (red arrows) must  occur.
The postshock molecular cloud fragments are likely to experience a distribution of shock velocities. Arrow number 4 represents the low velocity MHD shock  excitation of H$_2$ gas described in \S~\ref{subsec_MHDshocks}. A cyclic mass exchange between the cold molecular gas and the warm gas phases is also supported by the high value of the  $\rm H_2$ ortho to para ratio (\S~\ref{subsec_MHDshocks}).
More energetic shocks may dissociate the molecular gas (arrows number 5).
They are necessary to account for the low O~{\sc i} luminosity.
Even more energetic shocks can ionize the molecular gas (arrow number 6).
This would bring cold $\rm H_2$ directly into the H~{\sc ii} component.
Such shocks have been proposed as the ionization source of the clouds in the 
Magellanic Stream  by \citet{Bland-Hawthorn2007} based
on hydrodynamical simulations of their interaction with the Galactic halo plasma.

Turbulent  mixing of hot, warm and cold gas are an alternative mass input of H~{\sc ii} gas. Turbulent mixing layers have been described by
\citet{Begelman1990} and \citet{Slavin1993}.
They result from the shredding of the cloud gas into fragments that become too small to survive evaporation due to heat conduction. It is represented by the thin blue and green arrows to the left of our cartoon. Turbulent mixing produces intermediate temperature gas that is thermally unstable. This
gas cools back to produce H~{\sc ii} gas that enters the cycle (thin green arrow).
It is relevant for our scenario to note that cold gas in mixing layers preserves its dust content. It is only heated to a few $10^5\,$K, well below temperatures for which thermal sputtering becomes effective. Further, metals from the dust-free hot plasma  brought to cool are expected to accrete on dust when gas cools and condenses. Far-UV O{\sc vi} line observations are needed to quantify the mass flow rate through this path of the cycle and thereby the relative importance of photoionization (arrow 1), ionizing shocks (arrow 6), and turbulent mixing with hot gas
in sustaining the observed rate ($\simeq 100$~M$_{\odot}$~yr$^{-1}$) of recombining H~{\sc ii} gas.

Within this dynamical picture of the postshock gas, cold molecular gas is not necessarily a mass sink.
Molecular gas colder than $50~$K does not contribute to the H$_2$ emission.
The mass accumulated in this cold H$_2$ gas depends on the ratio between the timescale of mechanical energy dissipation  (on which gas is excited by shocks through one of the red arrows 4, 5 and 6) and the cooling timescale along the black arrow 3. 

\section{Why $\rm \bf H_2$ is a dominant coolant of the postshock gas?}
\label{sec_implications_dissipation}

This section sketches the energetics of the molecular gas in relation to its interaction with the background plasma. 
An efficient transfer of the bulk kinetic energy of the gas to
turbulent motions of much lower velocities within molecular clouds
is required to make H$_2$ a dominant coolant of the postshock gas. 
We consider that the bulk kinetic energy of the clouds drives the gas cycling through the warm/cold H~{\sc ii}, H~{\sc i} and H$_2$ states of multiphase fragmentary clouds and feeds the cloud turbulence. This hypothesis is supported by the fact that the mechanical energy is the main energy reservoir (\S~\ref{subsec_energy_reservoirs}). The energetics of the gas are discussed within this framework.

\subsection{Cloud dynamics and energy transfer}\label{subsec_cloud_dynamics}

In the collision, the clouds do not share the same dynamics as the lower density intercloud gas. The clouds keep their preshock momentum and move at high velocity with respect to the background plasma. The kinetic energy recorded in the  $900~\rm km\,s^{-1}$ H$_2$ line width  is dominated by the bulk motion of the H$_2$ clouds in the plasma rest frame. The turbulent velocities required within clouds to excite H$_2$ have a much smaller amplitude (relative fragment velocities up  to $20$~km$\, \rm s^{-1}$). The cloud turbulence needs to be continuously sustained to balance energy dissipation.
An efficient transfer of the bulk kinetic energy to low velocity turbulence of the  molecular gas is required to make H$_2$ a dominant coolant of the postshock gas.  Dissipation within either the hot plasma or warm atomic gas, leading to radiation at X-ray or optical wavelengths that is not efficiently absorbed by the molecular gas, has to be minor.

We are far from being able to describe how the energy transfer occurs. We just make qualitative statements that would need to be quantified with numerical simulations in future studies.
 Cloud motions are subject to drag from the background flow \citep{Murray2004}. This drag corresponds to an exchange of momentum and energy with the background plasma which drives turbulence into the clouds.

The fragmented, multiphase structure of the clouds is likely to be a main key of the energy transfer.
The dynamical interaction with the background plasma flow generates relative velocities between cloud fragments because their acceleration depends on their mass and density.
The relative motions between cloud fragments  can lead to collisions, which  results in the dissipation of  kinetic energy (\S~\ref{subsec_MHDshocks}).
This dissipation will occur preferentially in the molecular gas because it is the coldest component with the lowest sound speed.
The magnetic field is likely to be important because it weaves the cloud H$_2$
fragments to the lower density cloud gas which is more easily entrained by the background plasma flow.
The mass cycle between cloud gas states contributes to the energy transfer in two ways.
(1) Gas cooling transfers the turbulent energy of the warm
H~{\sc ii} and H~{\sc i} to the H$_2$ gas.  (2) The thermal instability
which occurs when the gas cools both from $10^6~$K and from $10^4~$K transfers
a significant fraction of the gas thermal energy to turbulent kinetic energy. Numerical simulations \citep{2003ApJ...591..238S, Audit2005} show that the thermal instability inhibits energy-loss through radiation from the warmer gas and feeds turbulence in the cooler gas. The turbulent velocities generated by the thermal instability are found to be commensurate with the sound speed in the warmer gas and are  thus supersonic for the colder gas.

\subsection{Gas energetics}\label{subsec_energetics}

The cloud turbulent energy and the dissipated power per H$_2$ molecule may be written as
\begin{equation}
E_T = \frac{3}{2} \, m_{\rm H_2} \, \sigma_T^2
\end{equation}
\begin{equation}
P_T = \frac{E_T}{t_T} = 0.24 \, \left( \frac{\sigma_T}{10 \rm ~km\, s^{-1}} \right)^2 \left(\frac{t_T}{\rm 10^5 \, \rm yr} \right)^{-1}~ \frac{\rm L_{\odot}}{M_{\odot}} ,
\end{equation}
where $m_{\rm H_2}$ is the mass of the H$_2$ molecule  and  $\sigma_T $ the cloud turbulent velocity dispersion.
The second equation introduces an effective dissipation timescale, $t_T$ which
is often written as the ratio between the cloud size, $L$, and $\sigma_T$. The reference value of $\sigma_T$
used in the second equation is commensurate with the shock velocities inferred from the fitting of the H$_2$ excitation diagram in \S~\ref{subsec_MHDshocks}.

If dissipation of turbulent energy  powers the H$_2$ emission, the energy dissipation rate must be at least equal to the power radiated by H$_2$.
Using the  H$_2$ emission and the warm H$_2$ mass in Table~\ref{table_budgets}, we estimate the luminosity to mass ratio of the warm H$_2$ gas in SQ to be $\rm 0.5~L_\odot/M_\odot$.  The dissipated power $P_T$ may be somewhat larger if it powers the dissociative and ionizing shocks represented by the arrows 5 and 6 in Fig.~\ref{Fig_mass_flows2}.
This calculation implies a dissipation timescale of  $\sim 5 \times 10^4~$yr which is not much higher than the cooling time of the lower excitation warm H$_2$ gas which accounts for the S(0) and S(1) line emission (\S~\ref{subsec_MassFlows}).
Cold H$_2$ gas would thus be  heated by turbulent dissipation on timescales comparable to the gas cooling time.
If this is right, the cold molecular gas mass should not be much larger than the warm H$_2$ mass inferred from the Spitzer observations.
The turbulence dissipation timescale is more than three
orders of magnitude smaller than the
ratio between the H$_2$  gas bulk kinetic energy
and H$_2$ luminosity (section~\ref{subsec_context}). Thereby, the dissipation of the collision energy must involve a large ($\gtrsim 1\,000$) number of cycles where H$_2$ gas fragments are accelerated and dissipate their kinetic energy in shocks.

\section{Future tests of the model and open questions}\label{sec_future_tests}

In this section we provide a few observational predictions of the model related to questions that remain open. For each point,  we discuss which observations may test these predictions.

\subsection{Dust Emission}

Dust is a key element in our scenario. It must be present in the postshock clouds  for $\rm H_2$ to form. \citet{Xu1999, 2003ApJ...595..665X} present results from infrared imaging observations carried out with the \textit{Infrared Space Observatory (ISO)}. Far-infrared (FIR) emission was detected in the shock region and \citet{2003ApJ...595..665X} proposed that it arises from large dust grains  in the hot X-ray emitting plasma.
Clearly, the \textit{Spitzer} detection of molecular gas opens a new perpective on the origin of this emission.
In our model we expect that some of the thermal dust emission comes from the molecular gas. Outside local UV enhancement associated with star formation seen in \textit{GALEX} observations \citep{2005ApJ...619L..95X}, dust emission should appear spatially correlated with the $\rm H_2$ emission. Given that the value of the radiation field in the shock is low  ($G \sim 1-2$~G$_0$ \S~\ref{subsec_UVfield}), we expect  rather faint dust emission in the SQ ridge. This accounts for the weakness of the PAH features in the \textit{Spitzer} spectra. 
Deep \textit{Spitzer} imaging observations have been undertaken and will allow us to test this prediction. 
Moreover, observations with the \textit{Herschel Space Telescope} will have the required angular resolution and sensitivity to detect the long wavelength emission of the dust in the SQ shock. 

\subsection{Mass Budget}

The \textit{Spitzer} $\rm H_2$ observations only trace the warm molecular gas ($T \gtrsim 150$~K). This warm gas could be physically connected to a more
massive reservoir of cold $\rm H_2$ gas. This is a major open question about the physical state of the molecular gas, as well as about the mass and energy budget.
In a forthcoming paper, we will report deep IRAM CO observations that set a low upper limit on the CO(1-0) and (2-1) line emission from the SQ shock area.
The physical state of the shocked gas is likely to be different from that of GMCs in galactic disks which are gravitationally bound. It is thus not clear that the standard CO emission to $\rm H_2$ mass
conversion factor applies.  \textit{Herschel} observations of the dust emission are needed to trace cold gas independently of the CO/$\rm H_2$ ratio.

\subsection{Shock Energetics}

\textit{Spitzer} observations of the low-excitation $\rm H_2$
rotational lines do not provide a complete budget of the kinetic
energy dissipation.  To constrain the present scenario, observations
of both higher and lower excitation lines are required. Rovibrational
$\rm H_2$ lines would probe dissipation in more energetic shocks than
those inferred from the H$_2$ pure rotational lines.  If the molecular
gas only experiences low-velocity MHD shocks, our model predicts a 1-0
S(1) line flux of $3.6 \times 10^{-19}$~W~m$^{-2}$ within the aperture
$\mathcal{A}$ (Table~\ref{table_results}). This is likely to be a
strict lower limit since the dynamical interaction between the gas
phases may well drive higher shock velocities
(\S~\ref{subsec_gal_shocks}).  Observations of the O{\sc i}~[63~$\mu$m]
line would provide constraints on the  physical state of
colder molecular gas.  The luminosity ratio of $\rm H_2$ rotational
lines to O{\sc i} depends on the shock velocity and the gas
density. For the SQ ridge, the observed $\rm H_2$ 0-0 S(1)
line flux within the area $\mathcal{A}$ is $2 \times 10^{-17}$~W~m$^{-2}$, and our model predicts a $\rm H_2$ 0-0~S(1) to O{\sc i}~[63~$\mu$m] luminosity ratio of 2.5.  The Integral Field Spectrometer of the PACS instrument onboard the
\textit{Herschel} satellite should be able to detect the O{\sc
  i}[63~$\mu$m] line in one hour of integration.  The spectral
resolution of these observation will be a factor $\sim 5$ higher than
\textit{Spitzer}. 
The O{\sc i}~[63~$\mu$m] line profiles will allow to compare the kinematics across the different temperatures of the gas. If the emission arises from multiphase clouds where warm and cold components are mixed down to small scales, their kinematics should be the same.

\subsection{Star Formation in the postshock gas}\label{subsec_starformation}

Many galaxy collisions and mergers are observed to trigger IR-luminous bursts of star formation.
However, the absence or weakness of spectroscopic signatures of photoionization (dust or ionized gas lines) in the center of the SQ ridge \citep{2003ApJ...595..665X}, where bright $\rm H_2$ emission was detected by \textit{Spitzer}, show that star formation does not occur in this region. 
Within the physical framework of dynamical interaction between gas phases described in Sect.~\ref{sec_cycle} and  \ref{sec_implications_dissipation}, these observations  suggest that most of the H$_2$ emission arises from molecular fragments that are not massive enough or not long-lived enough to collapse.  Star formation may also be quenched by continuous heating of the molecular gas and/or by the magnetic
pressure if the dense cloud fragments are not magnetically supercritical (\S~\ref{subsec_gal_shocks}).
Star-forming regions possibly associated with the galaxy collision are present away from the center of the shock \citep{Xu1999, Gao2000, Smith2001a, 2002A&A...394..823L, 2003ApJ...595..665X}.



In our model, star formation may arise from  the collapse of pre-existing GMCs (see \S~\ref{subsec_clouds_evolution}).
An alternative route to star formation may be the formation of gravitationally unstable postshock fragments. This second possibility has been quantified by numerical simulations that have introduced the concept of turbulence-regulated star formation \citep{Mac2004}.
 \citet{Krumholz2005} present an analytical description of the results of simulations, where the star formation rate depends on the Mach number of the turbulence within clouds and the ratio between their gravitational and turbulent kinetic energy. The star formation rate increases with the Mach number, but it is low even for high Mach numbers when the ratio between binding and turbulent energy is low. In the center of the SQ ridge, the star formation may be low because the average gas density is low (see Table~\ref{table_budgets}), even if the amount of turbulent energy is high. The low gas density and large-scale dynamical context (high-velocity dispersion) does not favor the formation of gravitationnally bound molecular clouds, such as galactic GMCs.

\section{Conclusions}\label{section_conclusion}

Spitzer observations led to the surprising detection of H$_2$ luminous
extragalactic sources whose mid-infrared spectra are dominated by molecular hydrogen line emission.
The absence or weakness of dust features or lines of ionized gas
suggests the presence of large quantities of warm molecular gas with no or very little star formation.
This work focuses on one of these H$_2$ luminous galaxies: the Stephan's  Quintet (SQ) galaxy-wide shock created by the collision between an intruding galaxy and a tidal arm at a relative speed $\rm \sim 1000\, km\, s^{-1}$. SQ observations place molecular gas in a new context where one has (1) to account for the presence of  H$_2$ in a galaxy halo, (2) to characterize its physical state and (3) to describe its role as a cooling agent of an energetic phase of galaxy interaction. 
The aim of the paper was to answer three main questions: \textit{(i) Why is
there $ \rm H_2 $ in the postshock gas~? (ii) How can we account for the $ \rm H_2 $ excitation~? (iii) Why is H$_2$ the dominant coolant~?} 
We summarize the main results of our work along these three points and present perspectives for future studies.

We have presented and quantified a scenario where H$_2$ forms out of shocked gas.
H$_2$ formation results from the density structure of the preshock gas.
The collision velocity, $\rm \sim 600\, km\, s^{-1}$ in the  centre of mass frame of the collision,
is the shock velocity in the low density ($n_{\rm H}~\lesssim~10^{-2}\rm~cm^{-3}$) volume filling gas.
This produces a $\sim 5\times  10^6 \,$K, dust-free, X-ray emitting plasma.
The pressure of this background plasma drives slower shocks into denser gas within clouds. Gas heated to temperatures smaller than $\sim 10^6 \,$K keeps its dust and has time to cool within the $5\times 10^6 \,$yr collision age. Because the postshock pressure is too high for the warm neutral medium to be thermally stable, the gas cools down to cold neutral medium temperatures, condenses and becomes molecular. We show that for a wide range of postshock pressures, as well as realistic preshock  cloud sizes and densities, the H$_2$ formation timescale is shorter than the collision age and also than the turbulent mixing timescale of the warm gas with the background plasma.


Observations show that the H$_2$ non-thermal kinetic energy
is the dominant energy reservoir. It is larger than the  thermal energy of the plasma.
We propose that the H$_2$ emission is powered by the dissipation of this kinetic  energy.
The H$_2$ excitation can be  reproduced by a combination of
low velocity ($\sim 5 \ \rm to \ 20$~km~s$^{-1}$) MHD shocks within dense ($n_{\rm H} \, > \, 10^3 ~\rm cm^{-3}$) H$_2$ gas.
In this interpretation, the pressure of the warm H$_2$ gas  is much higher than the ambient pressure set
by the background plasma. Such a pressure enhancement is
required to account for the higher excitation H$_2$ S(3) and S(5) rotational lines. For this shock velocity range, the warm H$_2$ pressure in the shocks is $\sim 0.5 - 5 \times 10^{8}$~K~cm$^{-3}$, $2-3$ orders of magnitude greater than the ambient pressure ($2 \times 10^{5}$~K~cm$^{-3}$).

The thermal, Rayleigh-Taylor and Kelvin-Helmhotz  instabilities
contribute to produce fragmentary postshock clouds where H$_2$, H{\sc i} and H{\sc ii} gas are mixed on small scales and embedded in the hot plasma.
In our scenario, the H$_2$ excitation is related to the dynamical interaction between gas phases which drives an energy transfer between these phases. 
An efficient transfer of the  kinetic energy associated with the bulk displacements of the clouds to turbulent motions of much lower velocities within molecular gas, is required to make H$_2$ a dominant coolant of the postshock gas.

We present a global view  at the postshock gas evolution that connects
H$_2$, warm H{\sc i} and H{\sc ii} gas emission to mass cycling between these gas components.
Within this dynamical picture of the postshock gas, cold molecular gas is not necessarily a mass sink. The fact that there is no star formation in the shocked region where H$_2$ is detected suggests that molecular fragments are not long-lived enough to collapse and become gravitationally unstable.

This global description provides a physical framework of general relevance for the interpretation of observational signatures, in particular H$_2$ emission, of mechanical energy dissipation in  multiphase gas. This study highlights the need to take into account H$_2$ formation and the mechanical energy dissipation within cold molecular gas in the physics of turbulent mixing of gas phases.


Future observations of the Stephan's Quintet will contribute to test the
proposed interpretation and better characterize some of its aspects. Some of these observations have been briefly discussed. Further work is needed to explore the impact of key physical parameters (gas pressure, radiation field, mechanical energy deposition, thermal to mechanical energy ratio, metallicity) on gas evolution  and on energy dissipation for the diversity of relevant galaxy-wide environments and mechanical energy sources (gas accretion, galaxy interaction and feedback from starburst and active galactic nuclei).
The analytical approach presented here should be pursued to compute how the timescales of the relevant physical processes depend on these parameters. The non-linear dynamical interaction between gas phases, in particular the driving of clouds turbulence, can only be investigated
quantitatively through numerical simulations.

\begin{acknowledgements}
	We thank Pierre-Alain Duc, Edith Falgarone, Patrick Hennebelle,
Mathieu Langer, Matthew Lehnert, Vincent Guillet, Pierre Lesaffre, Nicole Nesvadba and Patrick Ogle for discussions that helped us in our work. We appreciate the
helpful feedback from the referee, which guided us to improve the presentation
of our work.
\end{acknowledgements}

\bibliographystyle{aa.bst}
\bibliography{H2_SQ.bbl}

\Online

\begin{appendix}

\section{Modeling Dust Evolution}\label{appendix_dust_evol}

Interstellar dust plays a major role in the evolution of the postshock gas because its survival is required for $\rm H_2$ formation. For a constant dust-to-gas mass ratio, the dust dominates the cooling efficiency of the gas at high ($T> 10^{6}$~K) temperatures \citep{1987ApJ...322..812D} but for such high temperatures, the timescale for dust destruction is shorter than the gas cooling time  \citep{1996ApJ...473..864S}.
This appendix details how the evolution of the dust-to-gas mass ratio in the postshock gas has been calculated. This calculation is coupled to that of the gas cooling, detailed in Appendix~\ref{appendix_hotgascooling}.

The dominant dust destruction process depends on the shock velocity.
In this study we consider the effect of thermal and inertial sputtering by grain-ion collisions.  This is the dominant mode of destruction in a hot ($\sim 10^6$-$10^8$~K) plasma resulting from fast ($100 -1\,000$~km~s$^{-1}$) intergalactic shocks \citep[e.g.][]{1979ApJ...231...77D, 1996ApJ...457..244D}.

It is assumed that dust grains have an initial radius equal to the effective  (mean)  radius of $a_{\rm eff} = 0.05 \, \mu$m. This radius has been calculated from the MRN dust size distribution (\S~\ref{subsec_coolfunct_plasma}). At each time step, the fraction of dust remaining in the gas $f_{\rm dust}$, is computed in parallel to the postshock gas cooling:
\begin{equation}
f_{\rm dust} = \left( \frac{a}{a_{\rm eff}} \right) ^{3} \ \rm with \ a =  a_{\rm eff} - \int _{t_0} ^{t} \dot{a} \, dt \ ,
\end{equation}
where $ \dot{a} = da / dt$.
For a shock velocity $V_{\rm s} > 300$~km~s$^{-1}$, thermal sputtering dominates over inertial sputtering, so we ignore inertial sputtering. In this case the grain sputtering rate is nearly constant for  $T \gtrsim 3 \times 10^6 \,$K and strongly decreases at lower temperatures \citep{1979ApJ...231...77D, 1994ApJ...431..321T, 1996ApJ...457..244D}. The rate of decrease in grain size is given by
\begin{equation}
\frac{d a}{d t} = \frac{m_{\rm sp}}{2 \, \rho _0} \, n_{\rm H} \, \sum A_i \left\langle Y_i \, v \right\rangle \ ,
\end{equation}
where $m_{\rm sp}$ and $\rho _0$ are the average mass of the sputtered atoms  and the density of the grain material, $A_i$ the abundance of impacting ion $i$ and $ Y_i $ is the sputtering yield\footnote{The yield is the number of atoms ejected from the grains surface per incident particle.} of ion $i$. The quantity $\left\langle Y_i \, v \right\rangle$ is the average of the impacting ion velocities $v$ times the yield over the Maxwellian distribution. The right hand side of this equation is computed from the results given in \citet{1994ApJ...431..321T}.

For $V_{\rm s} < 300$~km~s$^{-1}$, we integrate the equation of deceleration of the dust grains with respect to the gas
\begin{equation}\label{eq_dust_decel}
\frac{d v_g}{d t}  = - \frac{\beta \, \pi \, a ^2 \, \rho \, v_{g}^{2}}{m} \ ,
\end{equation}
where $v_g$ is the grain velocity with respect to the gas, $m$ and $a$ are its mass and radius, $\rho$ is the gas density, and $\beta$ is the enhancement of the collisional drag in a plasma relative to that of a neutral medium. The initial grain velocity is set to $3/4$ of the shock velocity \citep{1994ApJ...433..797J}. We take $\beta =1$, which maximizes the dust destruction (see \citet{1995ApJ...454..254B} for details). Therefore, our computation of the dust survival is conservative.
When inertial sputtering is the dominant destruction process, the erosion rate is given by \citep{1994ApJ...431..321T}
\begin{equation}
\frac{d a}{d t} = \frac{m_{\rm sp}}{2 \, \rho _0} \, v_g \, n_{\rm H} \sum A_i  Y_i  \ .
\end{equation}
By integrating these differential equations, the radius of the dust grains as a function of their velocity has been deduced (see \citet{1995ApJ...454..254B} for a complete parametrization of $da/dt$ in the case of inertial sputtering).

%
\begin{figure}
   \includegraphics[angle=90, width = \columnwidth]{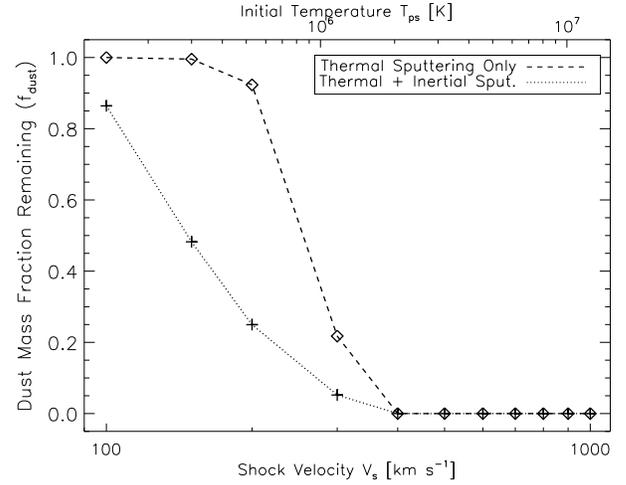}
   \caption{Dust mass fraction as a function of the initial (postshock) temperature, after isobaric cooling to $10^{4}\,$K. The pressure is fixed to $P_{\rm ps} / k_{\rm B} = 2.3 \times 10^{5} \ [\rm cm^{-3} \, K]$. The dotted line shows the result of the model which includes both inertial and thermal sputtering. As a comparison, the dashed line only shows the effect of thermal sputtering.}
   \label{Fig_DustMassFracRemain_InitTemp}
\end{figure}

Fig.~\ref{Fig_DustMassFracRemain_InitTemp} shows the dust mass fraction remaining in the gas after isobaric cooling from the postshock temperature $T_{\rm ps}$ to $10^{4}\,$K as a function of the shock velocity. The dotted line includes both thermal and inertial sputtering. For comparison, the dashed line only includes dust destruction by thermal sputtering. Our calculation for thermal sputtering with a single grain size ($a_{\rm eff} = 0.05 \, \mu$m) is in very good agreement with the study by \citet{1996ApJ...473..864S} who used  a grain size distribution. For $T_{\rm ps} > 2 \times 10^{6}\,$K, i.e. $V_{\rm s} > 400\,$km~s$^{-1}$, most of the dust is destroyed before the gas has cooled down to $10^{4}\,$K. The gas which is heated to less than $\sim 10^{6}\,$K keeps a major part of its dust content and may therefore form $\rm H_{2}$.

\section{Modeling the cooling of hot gas}\label{appendix_hotgascooling}

\begin{figure*}[t]
\begin{center}
 \includegraphics[height = \textwidth, angle=90]{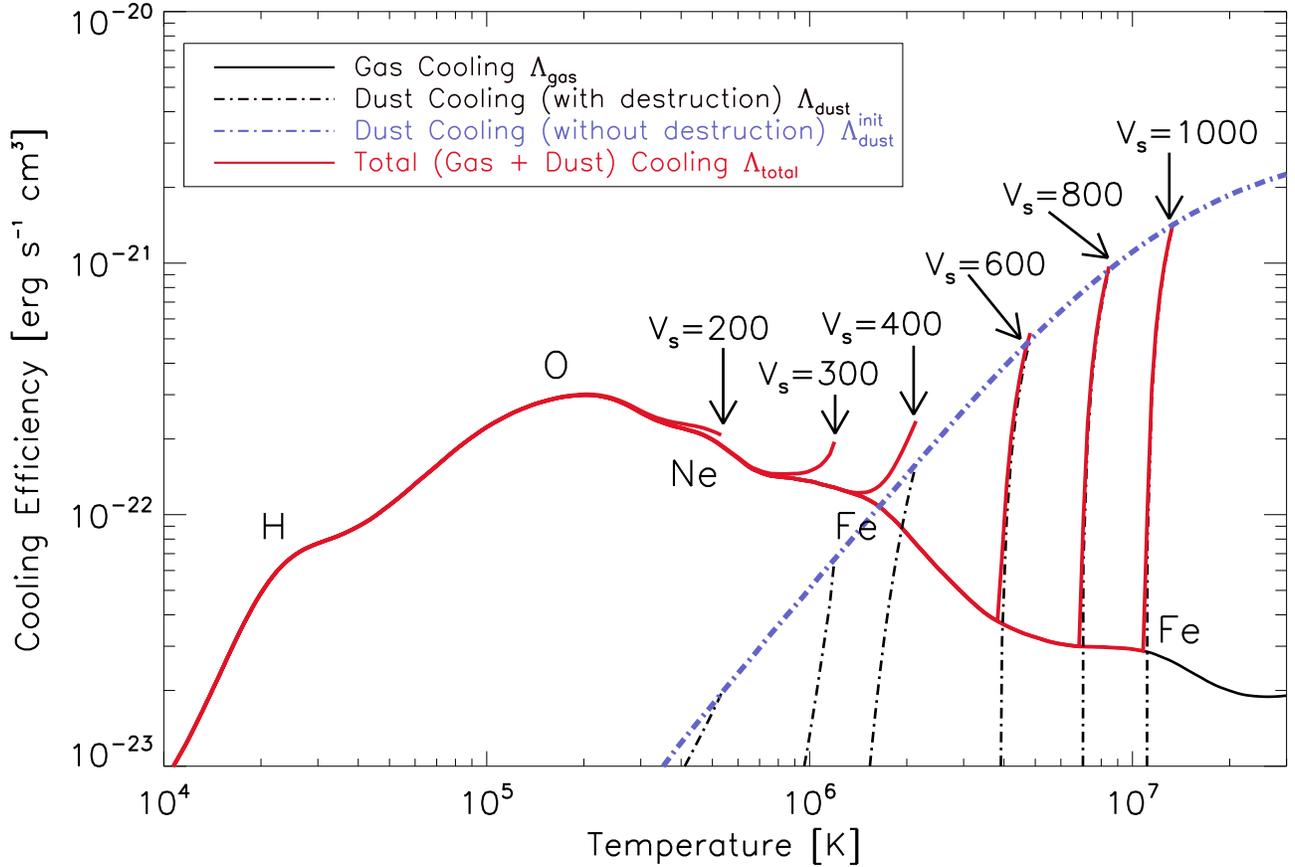}
 \caption{Time-dependent cooling efficiency [erg cm$^3$ s$^{-1}$] as a function of temperature during the gas cooling, for different initial conditions. The blue dashed line, $\Lambda _{\rm dust}^{\rm init} (T)$, represents the \emph{initial} cooling function of the gas including the dust contribution, computed for a MRN interstellar dust size distribution ($0.01$ to $0.25 \; \mu$m dust particles).
The black dashed lines are the dust cooling functions for different shock velocities  (i.e. different initial temperatures) which take into account the destruction by sputtering during the cooling.
The cooling function due to atomic processes, $\Lambda _{\rm gas} (T)$, is displayed for an isobaric and non-equilibrium (time-dependent) cooling, and for solar metalicities ($Z=1$). The dominant cooling elements at various temperatures are indicated on the curves. The red lines are the  total cooling functions, $\Lambda _{\rm total} = \Lambda _{\rm dust} + \Lambda _{\rm gas}$, for different initial temperatures, or shock velocities. The shock velocities are indicated in km~s$^{-1}$. Their starting points are indicated by the arrows.}
\label{Fig_cool_function}
\end{center}
  \end{figure*}

This appendix details how we model the cooling of the hot gas. 
The cooling functions used in the calculations are presented in \S~\ref{subsec_coolfunct_plasma}. Then the results are discussed in \S~\ref{subsec_evollowdensgas}. 

The time dependence of both the gas temperature and the dust-to-gas mass ratio has been calculated. We start from gas at an initial  postshock temperature $T_{\rm ps}$ with a galactic dust-to-gas ratio, a solar metallicity and assume equilibrium ionization.
From a range of postshock temperatures to $10^4$~K, the isobaric gas cooling is calculated by integrating the energy balance equation which gives the rate of decrease of the gas temperature:
\begin{equation}\label{Eq_ratetemp}
\frac{5}{2}  \, k_{\rm B} \, \frac{d T}{d t} = - \mu \, n_e \left(f_{\rm dust} \, \Lambda _{\rm dust} + \Lambda _{\rm gas}\right) \ ,
\end{equation}
where $\mu$ is the mean particle weight ($\mu = 0.6~\rm a.m.u$ for a fully ionized gas), $n_e$ is the electronic density, $k_{\rm B}$ the Boltzman constant, $f_{\rm dust}$ the dust-to-gas mass ratio, $\Lambda _{\rm dust}$ and $\Lambda _{\rm gas}$ are respectively the dust and gas cooling efficiencies per unit mass of dust and gas, respectively.

The time-dependent total cooling function is the sum of the dust cooling (weighted by the remaining fraction of dust mass, see Appendix~\ref{appendix_dust_evol}) and the gas cooling contributions. 
We neglect the effect of the magnetic field on the compression of the gas. This assumption is in agreement with observations and numerical simulations provided that the gas is not gravitationally bound (see \S~\ref{subsec_gal_shocks} for details and references). The thermal gas pressure $P_{\rm th}$ is constant in the calculations.




\subsection{Physical Processes and Cooling Functions}
\label{subsec_coolfunct_plasma}

\subsubsection{Cooling by atomic processes in the gas phase}

The calculations of the non-equilibrium (time-dependent) ionization states and isobaric cooling rates of a hot dust-free gas for the cooling efficiency \emph{by atomic processes} has been taken from \citet{2007ApJS..168..213G}. The electronic cooling efficiency, $\Lambda _{\rm gas}$, includes the removal of electron kinetic energy via recombinations with ions, collisional ionizations, collisional excitations followed by line emissions, and thermal bremsstrahlung. $\Lambda _{\rm gas}$ is shown on Fig.~\ref{Fig_cool_function} (black solid line\footnote{the black solid line (gas cooling efficiency) is merged with the total cooling curve (red line) for $T \lesssim 10^7$~K.})  and reproduces the standard ``cosmic cooling curve'' presented in many papers in the literature  \citep[e.g.][]{1993ApJS...88..253S}.
Note that most of the distinct features that appear for the ionization equilibrium case are smeared out in the nonequilibrium cooling functions and that the nonequilibrium abundances reduces the cooling efficiencies by factors $2$ to $4$ \citep{1993ApJS...88..253S, 2007ApJS..168..213G}.
The chemical species that dominate the gas cooling are indicated on Fig.~\ref{Fig_cool_function} near the $\Lambda _{\rm gas}$ cooling curve. Over most of the temperature range, the radiative energy losses are dominated by electron impact of resonance line transitions of metal ions. At $ T \simeq 2 \, 10^4 \,$K, the cooling is mainly due to the collisional excitation of hydrogen Ly$\alpha$ lines. For $T \gtrsim 10^7 \,$K, the bremsstrahlung (free-free) radiation becomes dominant.

\subsubsection{Cooling by dust grains}

Following the method of \citet{1987ApJ...322..812D}, the cooling function of the gas \emph{by the dust} via electron-grain collisions has been calculated. We adopt an MRN \citep{1977ApJ...217..425M} size distribution of dust particles, between $0.01$ to $0.25 \; \mu$m in grain radius. All the details can be found in \citet{1986ApJ...302..363D, 1987ApJ...322..812D}, and we briefly remind here the principles of this calculation. For temperatures smaller than $10^{8}$~K, electron - dust grain collisions cool the gas with a cooling efficiency, $\Lambda _{\rm dust} (T)$ $[\rm erg\, s^{-1} \, cm^{3}]$,  given by:
\begin{equation}
\Lambda ^{\rm init}_{\rm dust} (T) = \frac{\mu \, m_{\rm H} \, Z_{\rm d}}{\langle m_{\rm d} \rangle} \left( \frac{32}{\pi \, m_e} \right)^{1/2} \! \pi (k_{\rm B}  T)^{3/2} \! \int \! a^{2}  h(a, T)  f(a) \, da
\end{equation}
where $Z_{\rm d}$ is the dust-to-gas mass ratio, $m_{\rm H}$ the mass of an hydrogen atom, $m_e$ the mass of electron, $\mu$ the mean atomic weight of the gas (in amu), $f(a)$ the grain size distribution function (normalized to $1$) in the dust size interval, $\langle m_{\rm d} \rangle$ the size-averaged mass of the dust, and $h(a, T)$ the effective grain heating efficiency \citep[see e.g.][]{1981ApJ...248..138D, 1987ApJ...322..812D}. This \emph{initial} cooling efficiency is shown on the \emph{blue} dashed line of Fig.~\ref{Fig_cool_function} for a MRN dust size distribution and the solar neighbourhood dust-to-gas mass ratio of $7.5 \; 10^{-3}$.

The comparison on Fig.~\ref{Fig_cool_function} of the \emph{initial} cooling efficiency by the dust, $\Lambda _{\rm dust}^{^{\rm init}} (T)$,  and by atomic processes, $\Lambda _{\rm gas}$, shows that, for a dust-to-gas mass ratio of $Z_{\rm d} = 7.5 \; 10^{-3}$, the dust is initially the dominant coolant of the hot postshock gas for $T \gtrsim 10^6 \,$K. 
During the gas cooling, we calculate how much the grains are eroded and deduce the mass fraction of dust which remains in the cooling gas as a function of the temperature.  The total cooling function (\emph{red} curves on Fig.~\ref{Fig_cool_function}) are deduced by summing the gas and dust cooling rates.   We show different cooling efficiencies for different initial temperatures, corresponding to different shock velocities.

\subsection{Results: evolution of the hot gas temperature and dust survival}\label{subsec_evollowdensgas}

\begin{figure}
   \includegraphics[angle=90, width = \columnwidth]{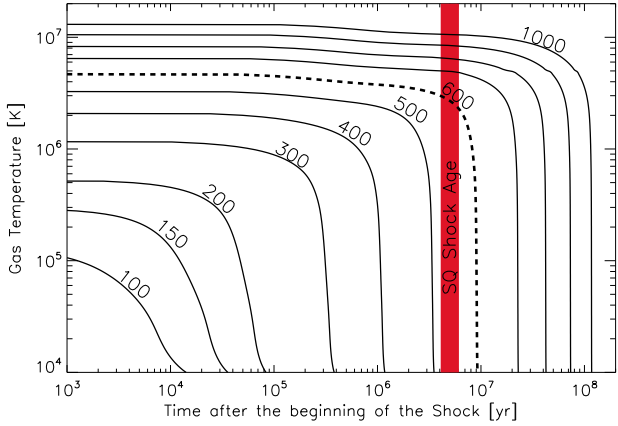}
   \caption{Evolution of the gas temperature with time at constant pressure, from the postshock temperature to $10^4$~K for different shock velocities (indicated on the curves in km s$^{-1}$). The dashed line is for $V_s = 600$~km~s$^{-1}$, which corresponds to the SQ hot ($5 \times 10^6$~K) plasma. The red vertical thick  line at $5 \times 10^6$~yr indicates the collision age.}
   \label{Fig_temp_profiles}%
\end{figure}

 The temperature profiles are shown on Fig.~\ref{Fig_temp_profiles} for different shock velocities ($100$ to $1\,000$~km~s$^{-1}$), which corresponds to a range of preshock densities of $\sim 0.2$ to $2 \times 10^{-3}$~cm$^{-3}$ for a postshock pressure of $2 \times 10^5$~K~cm~$^{-3}$.

\begin{figure}
   \includegraphics[angle=90, width = \columnwidth]{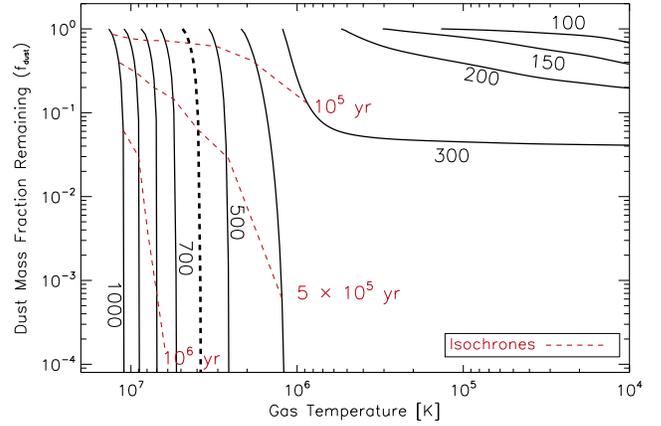}
   \caption{Dust mass fraction remaining as a function of the gas temperature, from the postshock temperature to $10^4$~K for different shock velocities $V_s$ [km s$^{-1}$] labelled on the curves. The gas cooling is isobaric, and the thermal gas pressure is set to the average pressure of the SQ hot gas $\displaystyle P_{\rm ps}/k_{\rm B} \simeq 2.3 \times 10^{5} \ \rm cm^{-3} \, K$. The dashed line is for $V_s = 600\,$km~s$^{-1}$. The red dashed lines are the cooling isochrones at $10^5$, $5\times 10^5$ and $10^6$~yr.}
   \label{Fig_dust_temp}%
\end{figure}

Fig.~\ref{Fig_dust_temp} shows the remaining fraction of dust mass as a function of the gas temperature. In this plot, the thermal gas pressure is constant and equals the measured average thermal gas pressure of the Stephan's Quintet hot gas  $\displaystyle P_{\rm ps}/k_{\rm B} \simeq 2 \times 10^{5} \ [\rm cm^{-3} \, K]$.
This plot illustrates a dichotomy in the evolution of the dust to gas mass ratio. On the top right side, a significant fraction of the dust is surviving in the gas, whereas on the left side, almost all the dust is destroyed on timescales shorter than the collision age ($5 \times 10^6$~yr).
In the intercloud gas shocked at high velocities ($V_{\rm s} > 400\,$km~s$^{-1}$), the dust mass fraction drops rapidly.
In clouds where the gas is shocked at $V_{\rm s} < 300\,$km~s$^{-1}$, the gas keeps a large fraction of its dust content ($> 20\,$\%).
For the gas which is heated to temperatures $T > 10^{6}$~K, the dust lifetime is smaller than the gas cooling time and most of the dust is destroyed.
At lower temperatures ($T < 10^{6}$~K), the dust cooling rate is smaller than the gas cooling rate.
Then, the dust never contributes significantly to the cooling of the postshock gas. This last result is in agreement with the earlier study by \citet{1996ApJ...473..864S}.

\section{Modeling $\rm H_2$ formation}
\label{appendix_H2}

\subsection{Cooling function for $T < 10^4$~K}
\label{subsec_lowT_chemistry}

This section describes the cooling of clouds below $10^4$~K and H$_2$ formation.
We use the chemistry, and the atomic and molecular cooling functions described in \citet{2003MNRAS.341...70F} and references therein. The principal coolants are O, $\rm H_2$, $\rm H_{2}O$ and OH \citep[see Fig.~3 in][]{2003MNRAS.341...70F}. 
The time evolution of the gas temperature and composition is computed at a fixed thermal gas pressure equal to that of the intercloud gas ($2 \times 10^{5}$~K~cm$^{-3}$). 
The metallicity and the gas-to-dust mass ratio are assumed to be the solar neighbourhood values. The initial ionization state of the gas is the out-of-equilibrium values resulting from cooling from higher temperatures \citep{2007ApJS..168..213G}. We assume a standard value for the cosmic ray ionization rate of $\zeta = 5 \times 10^{-17}$~s$^{-1}$ and the UV radiation field is not considered here (see \S~\ref{subsec_mass_nrj_budget}). The initial temperature is $10^4$~K and density $n_{\rm H} = 10$~cm$^{-3}$.
Hydrogen is initially highly ionized. During the postshock gas cooling, hydrogen recombination occurs, the neutral gas cools, $\rm H_2$ starts to form and further cools and condenses the gas.

\begin{figure}
 \includegraphics[width=\columnwidth]{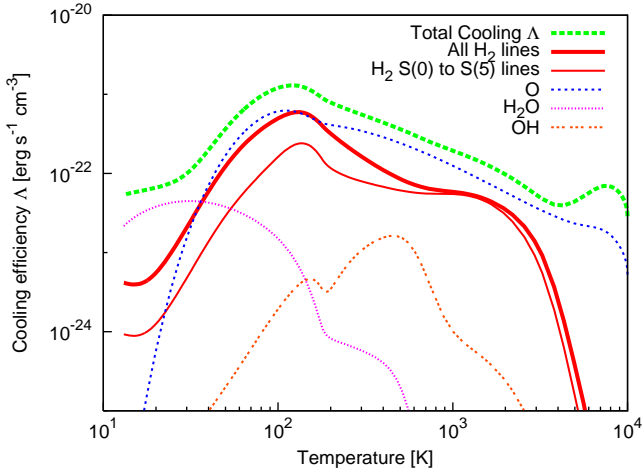}
 \caption{Cooling efficiency [erg~s$^{-1}$~cm$^{-3}$ by $\rm H_2$ S(0) to S(5) rotational lines (thin red line) and all $\rm H_2$ lines (thick red line) for the gas cooling at constant thermal pressure. Initially, the gas is ionized (see text for details), the density is $n_{\rm H} = 10$~cm$^{-3}$ and temperature is $T =10^4$~K. For comparison, the local cooling rate of all the coolants is shown (green dashed line).}
 \label{Fig_CoolFunct_isobar_vs_T}
\end{figure}

Fig.~\ref{Fig_CoolFunct_isobar_vs_T} presents the contribution of $\rm H_2$ line emission to the total cooling function as a function of the gas temperature. It shows the cooling functions of \emph{(i)} all the $\rm H_2$ lines (thick red line), \emph{(ii)} the pure $\rm H_2$ rotational lines S(0) to S(5) detected by \emph{Spitzer} (thin red line),  \emph{(iii)} the total cooling efficiency in which all elements are included (green dashed line).
The cooling efficiencies of some other major coolants (O in blue, $\rm H_{2}O$ in purple, and OH in orange) are also indicated.
H$_2$ excitation at low temperatures ($\sim 10$~K) is dominated by the contribution associated with $\rm H_2$ formation.

\subsection{$\rm H_2$ formation}\label{subsec_cloudcoolingH2form}

\begin{figure}
 \includegraphics[width=\columnwidth]{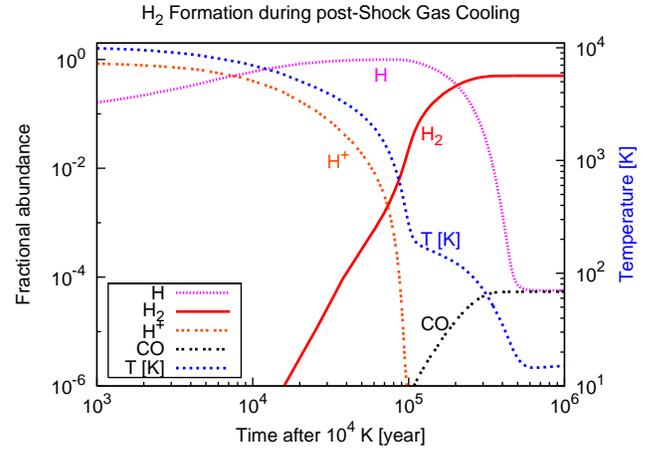}
 \caption{Isobaric cooling of the postshock gas and $\rm H_2$ formation. The temperature (blue line, labeled on the right side) and abundances profiles relative to $n_{\rm H}$ are shown. 
The initial conditions and model are the same than in Fig.~\ref{Fig_CoolFunct_isobar_vs_T}.
The final density of the molecular gas is $n(\rm H_2) = 2 \times 10^4$~cm$^{-3}$, for a temperature of $\sim 10$~K.}
 \label{Fig_species_time}%
\end{figure}

The chemical abundances profiles in the postshock gas are illustrated in Fig.~\ref{Fig_species_time} as a function of time, for the same model than in Fig.~\ref{Fig_CoolFunct_isobar_vs_T}
From the time when the gas attains $10^{4} \,$K, it takes $\sim 3 \times 10^{5}\,$yrs to form $\rm H_2$. This time scales inversely with the dust-to-gas mass ratio.
The $\rm H_2$ formation gives rise to a shoulder in the temperature profile around $200 \,$K. At this point, the energy released by the $\rm H_2$ formation is roughly
balanced by the cooling due to atomic Oxygen \citep[see Fig.~3 in][]{2003MNRAS.341...70F}.

\end{appendix}

\end{document}